\def\etal{{et al.\thinspace}}
\def\mearth{{\rm\,M_\oplus}}

\def\mdash{--}

\documentclass[apjl]{emulateapj}

\slugcomment{Icarus, accepted 5/20/2009}

\usepackage{marvosym}
\usepackage{ulem}

\received{} 
\accepted{} 

\lefthead{} \righthead{}

\begin{document}

\shorttitle{Building the Terrestrial Planets} 
\shortauthors{Raymond, O'Brien, Morbidelli, \& Kaib}

\title{Building the Terrestrial Planets: Constrained Accretion in the Inner Solar System}

\author{Sean N. Raymond\altaffilmark{1}, David P. O'Brien\altaffilmark{2},
Alessandro Morbidelli\altaffilmark{3}, \& Nathan A. Kaib\altaffilmark{4}}

\altaffiltext{1}{Center for Astrophysics and Space Astronomy, University of
Colorado, UCB 389, Boulder CO 80309-0389; rayray.sean@gmail.com}
\altaffiltext{2}{Planetary Science Institute, Tucson, AZ}
\altaffiltext{3}{Observatoire de la C\^ote d'Azur, Boulevard de l'Observatoire, BP 4229, 06304 Nice Cedex 4, France.} 
\altaffiltext{4}{Department of Astronomy,
University of Washington, Seattle, WA 98195}

\begin{abstract}
To date, no accretion model has succeeded in reproducing all observed
constraints in the inner Solar System.  These constraints include 1) the
orbits, in particular the small eccentricities, and 2) the masses of the
terrestrial planets -- Mars' relatively small mass in particular has not been
adequately reproduced in previous simulations; 3) the formation timescales of
Earth and Mars, as interpreted from Hf/W isotopes; 4) the bulk structure of the
asteroid belt, in particular the lack of an imprint of planetary embryo-sized
objects; and 5) Earth's relatively large water content, assuming that it was
delivered in the form of water-rich primitive asteroidal material.  Here we
present results of 40 high-resolution (N=1000-2000) dynamical simulations of
late-stage planetary accretion with the goal of reproducing these constraints,
although neglecting the planet Mercury.  We assume that Jupiter and Saturn are
fully-formed at the start of each simulation, and test orbital configurations
that are both consistent with and contrary to the ``Nice model.''  We find
that a configuration with Jupiter and Saturn on circular orbits forms
low-eccentricity terrestrial planets and a water-rich Earth on the correct
timescale, but Mars' mass is too large by a factor of 5-10 and embryos are
often stranded in the asteroid belt.  A configuration with Jupiter and Saturn
in their current locations but with slightly higher initial eccentricities ($e
= 0.07-0.1$) produces a small Mars, an embryo-free asteroid belt, and a reasonable
Earth analog but rarely allows water delivery to Earth.  None of the
configurations we tested reproduced all the observed constraints.  Our
simulations leave us with a problem: we can reasonably satisfy the observed
constraints (except for Earth's water) with a configuration of Jupiter and
Saturn that is at best marginally consistent with models of the outer Solar
System, as it does not allow for any outer planet migration after a few Myr.
Alternately, giant planet configurations which are consistent with the Nice
model fail to reproduce Mars' small size.
\end{abstract}


\section{Introduction}

It is commonly accepted that rocky planets form by the process of collisional
agglomeration of smaller bodies (for recent reviews, see Chambers 2004,
Nagasawa \etal 2007 or Raymond 2008).  This process starts from micron-sized
dust grains in young circumstellar disks, and the current paradigm proceeds as
follows.  Grains settle to a thin disk midplane on a $\sim 10^4$ year
timescale (Weidenschilling 1980), and grow quickly via sticky collisions until
they reach cm- or m- sizes (Dullemond \& Dominik 2004).  The time for m-sized
bodies to spiral in to the star is very short ($\sim$ 100 years) such that
this size range constitutes a barrier to further growth (Weidenschilling
1977a).  This barrier may be crossed by rapid accretion (Weidenschilling \&
Cuzzi 1993; Benz 2000) or by local gravitational instability (Goldreich \&
Ward 1973; Youdin \& Shu 2002), which can be triggered by turbulent
concentration (Johansen \etal 2007; Cuzzi \etal 2008).  Larger bodies (100 m
to 100 km in size), which are more weakly coupled to the gaseous disk, are
called planetesimals.  Runaway growth of the largest planetesimals may occur
while the velocity dispersion is small because of strong gravitational
focusing such that $dM/dt \sim M^{4/3}$ (Safronov 1969; Greenberg \etal 1978).
However, viscous stirring by the large bodies increases the velocity
dispersion of planetesimals, thereby reducing the growth rate to a roughly
geometrical regime, where $dM/dt \sim M^{2/3}$ (Ida \& Makino 1993). Dynamical
friction acts on the oligarchs, maintaining small eccentricities (Ida \&
Makino 1992; Kokubo \& Ida 1998).  The building blocks of the terrestrial
planets, $\sim$Moon-sized planetary embryos, form in $10^5-10^6$ years with a
characteristic spacing of 5-10 mutual Hill radii (Wetherill \& Stewart 1993;
Weidenschilling \etal 1997; Kokubo \& Ida 2000, 2002).  Giant collisions
between planetary embryos begin to occur when the local density of
planetesimals and embryos is comparable (Wetherill 1985; Kenyon \& Bromley
2006).  During late-stage accretion, embryo-planetesimal and embryo-embryo
impacts are common and the feeding zones of terrestrial planets can span
several AU in width (Raymond \etal 2006a).  Late-stage accretion lasts for
$\sim 10^8$ years and sets the final bulk architecture of the system as well
as the composition of the terrestrial planets (e.g., Wetherill 1996).

Past simulations of late-stage accretion have succeeded in reproducing several
aspects of the Solar System's terrestrial planets.  Using only 20-165
particles, Agnor \etal (1999) and Chambers (2001) roughly reproduced the
approximate masses and semimajor axes of Mercury, Venus, Earth and Mars.
Thommes \etal (2008) also reproduced the rough mass distribution of the inner
Solar System by invoking sweeping secular resonances during the depletion of
the Solar Nebula. By taking dynamical friction from remnant planetesimals into
account, O'Brien \etal (2006) and Morishima \etal (2008) reproduced the very
low eccentricities of the terrestrial planets.  Several groups have succeeded
in delivering water to Earth from hydrated asteroidal material, following the
model of Morbidelli \etal (2000; see also Raymond \etal 2004, 2006a, 2007;
O'Brien \etal 2006).

Despite these achievements, no previous study has adequately reproduced all
aspects of the inner Solar System.  Indeed, as pointed out by Wetherill
(1991), Mars' small size remains the most difficult constraint to reproduce
(also discussed in Chambers 2001).  Agnor \etal (1999), Chambers (2001) and
Morishima \etal (2008) succeeded in reproducing Mars' small size only because
their simulations started from an annulus of material with a fixed width (see
also Kominami \& Ida 2002).  In most cases this annulus extended from 0.5-1.5
AU, such that a small planet could form at the outer edge of the initial disk
because of spreading.  However, no such edge is thought to have existed in the
Solar Nebula, so that the assumption that the planetesimal and embryo
population extended only to 1.5 AU is not justified. Chambers (2001) managed
to place Mercury within a planetary mass distribution but only by adopting an
ad-hoc inner disk profile.  Thommes \etal (2008) formed a small Mars but the
orbits they assumed for Jupiter and Saturn are inconsistent with any
significant late, planetesimal-driven migration of the giant planets
(discussed at length in \S 6.2 below).  In fact, the scenario of Thommes \etal
(2008) is incompatible with the two currently viable theories for the late
heavy bombardments because these require either a more compact configuration
of Jupiter and Saturn (Gomes \etal 2005) or the formation of a small,
sub-Mars-sized planet at $\sim$ 2 AU (Chambers 2007).

Terrestrial accretion lasts for $\sim 10^8$ years, far longer than the few Myr
lifetimes of the gaseous component of protoplanetary disks (Haisch \etal 2001;
Brice\~no \etal 2001; Pascucci \etal 2006).  Thus, gas giant planets must be
fully-formed during late-stage accretion and can therefore strongly affect
terrestrial bodies, especially if the giant planets' orbits are eccentric
(Wetherill 1996; Chambers \& Cassen 2002; Levison \& Agnor 2003; Raymond \etal
2004; O'Brien \etal 2006).  Given that substantial orbital migration of the
Solar System's giant planets has been proposed to explain the structure of the
Kuiper Belt (Fernandez \& Ip 1984; Malhotra 1995) and the origin of the late
heavy bombardment (Strom \etal 2005; Gomes \etal 2005), the orbits of Jupiter
and Saturn at early times are unclear.  Indeed, a range of Jupiter-Saturn
configurations could yield the current Solar System.  Thus, if any particular
configuration were especially adept at reproducing the terrestrial planets, it
would provide strong circumstantial evidence in favor of that configuration.

In this paper we attempt to reproduce the inner Solar System with a suite of
high-resolution (N = 1000-2000) dynamical simulations of late-stage accretion.
We only vary one parameter of consequence: the configuration of Jupiter and
Saturn at early times.  We quantify five relevant constraints that we use to
test our models in section 2.  In section 3, we outline our choices of initial
conditions and numerical methods.  In section 4 we explore the case of two
contrasting simulations that each reproduce certain constraints.  We present
results and analysis of all simulations in section 5.  We discuss these results
and present our conclusions in section 6.

\section{Inner Solar System Constraints}

We consider five broad attributes which we attempt to reproduce statistically
with accretion simulations.  Other observations and measurements exist for
inner Solar System bodies, but we are limiting ourselves to relatively broad
and well-understood characteristics.  These constraints are described below in
order from strongest to weakest.  Weaker constraints rely on models or data
that are subject to interpretation, while strong constraints are directly
observed.  We use several quantities to compare our simulations with the Solar
System's terrestrial planets.  These include statistical measures that were
introduced by Chambers (2001).

\begin{enumerate} 
\item The masses and the mass distribution of the terrestrial planets.  As
mentioned above, the mass distribution of the terrestrial planets, and in
particular the small masses of Mercury and Mars, have not been adequately
reproduced in the context of the entire Solar System and its history.  In this
paper we do not attempt to reproduce Mercury because its small size and large
iron content may be the result of a mantle-stripping impact (Benz \etal 1988)
or interesting composition-sorting gaseous effects (Weidenschilling 1978).
However, for the case of Mars, with its more distant orbit, these effects are
less likely to be a factor, and it should be reproducible in the context of
our simulations.  In addition, the distribution of mass in the inner Solar
System is interesting because the majority is concentrated between the orbits
of Venus and Earth.  We therefore use two statistical measures for this
constraint: 

\begin{itemize}
\item The number of planets formed $N_p$.  We take $N_p$ to represent objects
that contain at least one planetary embryo, that have semimajor axes $a <$ 2
AU, and that are on long-term stable orbits.  It is only for these planets
that we apply our other measures.

\item A radial mass concentration statistic $RMC$ (called $S_c$ in Chambers
2001): \begin{equation} RMC = max \left(\frac{\sum M_j}{\sum M_j
[log_{10}(a/a_j)]^2} \right), \end{equation} \noindent where $M_j$ and $a_j$
are the masses and semimajor axes of each planet.  The function in brackets is
calculated for $a$ throughout the terrestrial planet zone, and $S_c$ is the
maximum of that function.  This quantity represents the degree to which mass
is concentrated in a small radial annulus: $S_c$ remains small for a system of
many equal-mass planets but $S_c$ is large for systems with few planets and
with most of the mass in one or two planets.  For a one planet system, the
$RMC$ value is infinite.  The $RMC$ of the Solar System's terrestrial planets
is 89.9 (see Table 2).
\end{itemize}

\item The orbits of the terrestrial planets.  The terrestrial planets maintain
very small orbital eccentricities and inclinations over long timescales.
Earth and Venus' time-averaged eccentricities are only about 0.03 (e.g., Quinn
\etal 1991).  Recent simulations with $N \geq 1000$ particles have succeeded
in reproducing these small eccentricities for the first time (O'Brien \etal
2006).  We quantify the orbital excitation of the terrestrial planets using
the normalized angular momentum deficit $AMD$ (Laskar 1997).  This measures
the difference in angular momentum of a set of orbits from coplanar, circular
orbits: 
\begin{equation} 
AMD = \frac{\sum_{j} m_j \sqrt{a_j} \left(1 - cos
(i_j) \sqrt{1-e_j^2}\right)} {\sum_j m_j \sqrt{a_j}}, 
\end{equation}
\noindent where $a_j$, $e_j$, $i_j$, and $m_j$ refer to planet $j$'s semimajor
axis, eccentricity, inclination with respect to a fiducial plane, and mass.
The $AMD$ of the Solar System's terrestrial planets is 0.0018 (see Table 2).

\item The formation timescales of Earth and Mars.  Recent interpretation of
Hf/W measurements suggest that the last core-formation event on Earth occurred
at roughly 50-150 Myr (Touboul \etal 2007)\footnote{Touboul \etal's (2007)
core-formation age is roughly a factor of two longer than previous estimates
(Kleine \etal 2002; Yin \etal 2002).  It is important to note that Hf/W
measurements of Earth samples are somewhat uncertain given the unknown amount
of core/mantle equilibration during giant impacts (Halliday 2004; Nimmo \&
Agnor 2006).  However, the samples from Touboul \etal (2007) are lunar in
origin and therefore circumvent the issue of equilibration.}  This event is
thought to be the Moon-forming impact (Benz \etal 1986; Canup \& Asphaug
2001).  Mars' formation time from Hf/W isotopes appears to be significantly
shorter, about 1-10 Myr (Nimmo \& Kleine 2007).

\item The large-scale structure of the asteroid belt.  The asteroid belt shows
a clear division between inner, S-types and more distant C-types (e.g., Gradie
\& Tedesco 1982).  In addition, there are no large gaps in the main belt
except those caused by specific mean motion or secular resonances.  If a
planetary embryo above a critical mass were stranded in the asteroid belt for
a long period of time, it would disrupt both of these observed characteristics
by planetesimal scattering (O'Brien \etal, in preparation).  This constraint
puts an upper limit of a few lunar masses ($\sim 0.05 \mearth$) on the mass of
an object that can survive in the asteroid belt after terrestrial planet
formation.  If an embryo did end up in the main belt, it could have been
subsequently removed during the late heavy bombardment (Gomes \etal 2005;
Strom \etal 2005), but the embryo's dynamical imprint on the asteroid belt
would have remained.\footnote{ It is important to note that the late heavy
bombardment was a purely dynamical event, as shown by the difference between
crater size distributions on surfaces older vs. younger than 3.8 Gyr (Strom
\etal 2005).}  We note that the asteroid belt is thought have been depleted by
a factor of $\sim 10^4$ in mass over the lifetime of the Solar System.  This
depletion is best explained by scattering of planetesimals by planetary
embryos in the primordial belt (Wetherill 1992; Chambers \& Wetherill 2001;
Petit \etal 2001; O'Brien \etal 2007), although other models do exist (e.g.,
Lecar \& Franklin 1997; Nagasawa \etal 2000).  Scattering among embryos often
places one body in an unstable mean motion resonances with Jupiter, leading to
their rapid removal from the belt.  This scattering also leads to some radial
mixing, consistent with the observation that the different asteroid taxonomic
types are not confined to narrow zones, but are spread somewhat in overlapping
but still distinct regions (Gradie \& Tedesco 1982).  Embryos as small as the
Moon are able to provide the necessary excitation (Chambers \& Wetherill
2001).  Most of the embryos are removed on a timescale of $\sim$ 10 Myr.
However, if one or more stray embryos with too large of a mass remain in the
belt for much longer than this, they will lead to excessive radial mixing,
inconsistent with the observed distribution of different asteroid taxonomic
types.  Figure~\ref{fig:astemb} shows the effect of a Mars-mass embryo trapped
at 2.5 AU on 100 Myr of evolution of 1000 asteroids in the main belt (2-3.5
AU), which are assumed to be massless.  Two features from
Fig.~\ref{fig:astemb} are inconsistent with the observed main belt: the excess
radial mixing and the gap created in the vicinity of the embryo.  More massive
or more eccentric asteroidal embryos can be significantly more disruptive than
the case from Fig.~\ref{fig:astemb}, especially if their eccentricity is
strongly forced by secular perturbations from Jupiter and Saturn (O'Brien
\etal, in preparation).  In addition, the simulation from
Fig.~\ref{fig:astemb} was only run for 100 Myr, roughly 500 Myr shorter than
the relevant timescale, i.e., the time between the completion of terrestrial
accretion ($\sim$ 100 Myr) and the time of the late heavy bombardment (600-700
Myr).  Thus, the constraint we place on our accretion simulations is that no
embryos larger than 0.05 $\mearth$ can survive in the main belt past the end
of terrestrial planet growth, or in our case $2 \times 10^8$ years.

\begin{figure}
\centerline{\epsscale{1.}\plotone{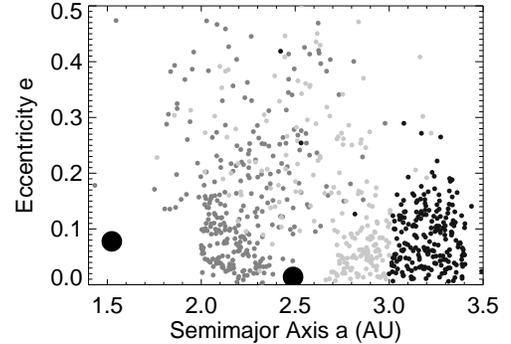}}
\caption{The effect of a Mars-sized planetary embryo on the structure of the
asteroid belt.  Shown are the surviving (massless) asteroidal bodies, whose
orbits were integrated for 100 Myr under the influence of Jupiter and Saturn
(not shown), Mars and a Mars-mass planetary embryo stranded in the asteroid
belt at 2.49 AU.  Asteroids are color-coded according to their starting
semimajor axes: grey (2-2.5 AU), light grey (2.5-3 AU), and black (3-3.5 AU).}
\label{fig:astemb}
\end{figure}

\item Earth's water content.  One prominent model suggests that primitive
asteroidal material was the source of the bulk of Earth's water (Morbidelli
\etal 2000; see also Raymond \etal 2007).  This model explains why the D/H
ratio of Earth's water matches that of carbonaceous chondrites (Robert \&
Epstein 1982; Kerridge 1985), and links Earth's water to the depletion of the
primitive asteroid belt.  Note that other models exist which propose that
Earth's water came from comets (Delsemme 1994; Owen \& Bar-Nun 1995), from
oxidation of a primitive, H-rich atmosphere (Ikoma \& Genda 2006), from
adsorption of water onto small grains at 1 AU (Muralidharan \etal 2008), or
from other sources -- see Morbidelli \etal (2000) for a discussion of some of
these models.  However, it is our opinion that the asteroidal water model of
Morbidelli \etal (2000) is the most likely source of Earth's water.  In fact,
water vapor from sublimation of in-spiraling icy bodies has been detected
interior to 1 AU in the protoplanetary disk around the young star MWC480
(Eisner 2007); this may be an observation of asteroidal (or in this case
potentially cometary) water delivery in action.

\end{enumerate}

\section{Methods}

Our simulations are designed to start at the beginning of late-stage
accretion, after Jupiter and Saturn are fully-formed and the nebular gas has
dissipated.  This is probably 1-3 Myr after ``time zero,'' and we base our
initial conditions on models of the formation of planetary embryos (e.g.,
Kokubo \& Ida 2000).  We start with a disk of planetary embryos and
planetesimals, plus Jupiter and Saturn.  Our simulations are comparable to the
highest-resolution cases in the literature, containing 85-90 planetary embryos
and 1000-2000 planetesimals.\footnote{The highest-resolution published
late-stage accretion simulations to date had N=2000-3000 (Raymond \etal 2006a;
Morishima \etal 2008).}

\subsection{Configuration of Jupiter and Saturn}

The resonant structure of the Kuiper Belt appears to require a significant
outward migration of Neptune (Fernandez \& Ip 1984; Malhotra 1995; Gomes 2003;
Levison \& Morbidelli 2003).  This outward migration occurred because of the
back-reaction from planetesimal scattering, which causes the orbits of Saturn,
Uranus and Neptune to expand and the orbit of Jupiter to contract (Fernandez
\& Ip 1984).  In addition, the ``Nice model'' of giant planet evolution, which
explains several observed characteristics of the Solar System, proposes that
Jupiter and Saturn formed interior to their mutual 2:1 mean motion resonance,
perhaps in fact in the 3:2 resonance and migrated apart (Tsiganis \etal 2005;
Gomes \etal 2005; Morbidelli \etal 2005, 2007).  Thus, Jupiter and Saturn may
very well have been in a more compact configuration at early times.

We tested a range of configurations for Jupiter and Saturn, although we did
not perform an exhaustive search given the large computational expense of each
simulation.  However, to account for stochastic variations in outcome we
performed 4 simulations for each giant planet configuration.  The
configurations we tested were: \begin{itemize}

\item CJS (``Circular Jupiter and Saturn'').  These are the initial conditions
for the Nice model, as in Tsiganis \etal (2005) and also used in O'Brien \etal
(2006).  Jupiter and Saturn were placed on circular orbits with semimajor axes
of 5.45 and 8.18 AU and a mutual inclination of 0.5 degrees.  We note that
even though Jupiter and Saturn begin with zero eccentricities, they induce
small, non-zero eccentricities in each others' orbits.

\item CJSECC.  Jupiter and Saturn were placed at their CJS semimajor axes of
5.45 and 8.18 AU with $e_J =$ 0.02 and $e_S =$ 0.03 and a mutual inclination of
0.5 degrees.

\item EJS (``Eccentric Jupiter and Saturn'').  Jupiter and Saturn were placed
on approximately their current orbits: $a_J$ = 5.25 AU, $e_J$ = 0.05, $a_S$ =
9.54 AU, and $e_S$ = 0.06, with a mutual inclination of 1.5 degrees.

\item EEJS (``Extra Eccentric Jupiter and Saturn'').  Jupiter and Saturn were
placed at their current semimajor axes but with higher orbital eccentricities:
$a_J$ = 5.25 AU, $a_S$ = 9.54 AU, and $e_J = e_S$ = 0.1, with a mutual
inclination of 1.5 degrees.  These cases proved to be interesting, so we ran 8
cases in addition to the original four.  The next four cases (referred to as
EEJS 5-8) had the same configuration of Jupiter and Saturn but 2000
planetesimals rather than 1000.  The final four cases (EEJS 9-12) also had
2000 planetesimals but had $e_J =$ 0.07 and $e_S =$ 0.08.

\item JSRES (``Jupiter and Saturn in RESonance'').  Jupiter and Saturn were
placed in their mutual 3:2 mean motion resonance, following directly from
simulations of their evolution in the gaseous Solar Nebula (Morbidelli \etal
2007): $a_J$ = 5.43 AU, $a_S$ = 7.30 AU, $e_J$ = 0.005, and $e_S$ = 0.01, with
a mutual inclination of 0.2 degrees.

\item JSRESECC (``Jupiter and Saturn in RESonance on ECCentric orbits'').  As
for JSRES but with $e_J = e_S$ = 0.03. \end{itemize}

The EJS and EEJS simulations assume that Jupiter and Saturn did not undergo
any migration.  The EEJS simulations are more self-consistent than the EJS
simulations, because scattering of remnant planetesimals and embryos tends to
decrease the eccentricities and semimajor axes of Jupiter and Saturn (e.g.,
Chambers 2001).  Thus, to end up on their current orbits, Jupiter and Saturn
would have had to form on more eccentric and slightly more distant orbits. The
CJS, JSRES and JSRESECC simulations all follow from the Nice model and assume
that Jupiter and Saturn's orbits changed significantly after their formation,
with Saturn migrating outward and Jupiter inward (Tsiganis \etal 2005).  If
migration of the giant planets is really associated with the late heavy
bombardment (Gomes \etal 2005; Strom \etal 2005), then at least most of the
migration of Jupiter and Saturn must have occurred late, well after the
completion of the terrestrial planet formation process.

\subsection{Properties of the Protoplanetary Disk}

For all of our simulations, the disk of solids extended from 0.5 to 4.5 AU and
contained populations of planetary embryos and planetesimals.  For most cases,
we assumed that the disk's surface density in solids $\Sigma$ followed a
simple radial power-law distribution:
\begin{equation} 
\Sigma(r) = \Sigma_1
\left(\frac{r}{\rm 1 AU}\right)^{-x}. 
\end{equation} 
\noindent For the minimum-mass solar nebula (MMSN) model, $\Sigma_1 \approx
6-7\,\, g\, cm^{-2}$ and $x=3/2$ (Weidenschilling 1977a; Hayashi 1981).  For
most of our simulations we assumed $x=3/2$ but we also performed some cases
with $x=1$ for the CJS and EJS giant planet configuration.  Cases with $x=1$
are labeled by the $x$ value; for example, the EJS15 simulations have $x=3/2$
and the EJS1 simulations have $x=1$ (see Table 2).  For each case, we
calibrated our disks to contain a total of 5 $\mearth$ in solids between 0.5
and 4.5 AU, divided equally between the planetesimal and embryo components.

\begin{figure}
\centerline{\epsscale{1.}\plotone{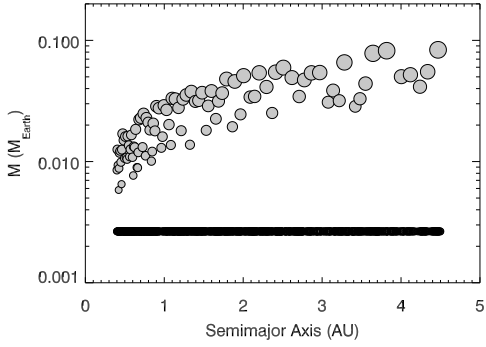}}
\caption{Sample initial conditions for a disk with $\Sigma \sim r^{-3/2}$
containing 97 planetary embryos and 1000 planetesimals.  Embryos are shown in
gray with their sizes proportional to their mass$^{(1/3)}$ (but not to scale
on the x axis).}
\label{fig:init}
\end{figure}

Figure~\ref{fig:init} shows a sample set of initial conditions.  We assumed
that embryos are spaced by $\Delta = $ 3-6 mutual Hill radii $R_H$, where $R_H
= 0.5 \, (r_1+r_2) \, [(M_1+M_2)/3 M_\odot]^{1/3}$, where $a_1$ and $M_1$ are
the radial distance and mass of embryo 1.  The embryo mass therefore scales
with orbital distance as $M \sim r^{3/2\,(2-x)}\Delta^{3/2}$ (Kokubo \& Ida
2002; Raymond \etal 2005).  The disks contained 85-90 embryos with masses
between 0.005 and 0.1 $\mearth$.  In Mars' vicinity the typical embryo mass
was roughly 1/6 to 1/3 of a Mars mass.  Planetesimals were laid out as $N_p
\sim r^{x+1}$ to follow the annular mass, and had masses of 0.0025
$\mearth$.  Embryos and planetesimals were given randomly-chosen starting
eccentricities of less than 0.02 and inclinations of less than 0.5$^\circ$. In
a few EEJS cases we performed additional simulations with 2000 planetesimals,
which followed the same distribution but had correspondingly smaller
masses. 

We assume that there existed a radial compositional gradient for rocky bodies
in the Solar Nebula.  This gradient was presumably imprinted on planetesimals
by the local temperature during their formation (e.g., Boss 1998), although
heating by short-lived radionuclides such as $^{26}$Al may have played a role
(Grimm \& McSween 1993).  We assume the same water distribution as in Raymond
\etal (2004, 2006a), using data for primitive meteorites from Abe \etal
(2000). The ``water mass fraction'', $WMF$, i.e. the water content by mass,
varies with radial distance $r$ as: 
\begin{equation} 
WMF = \left\{ \begin{array}{ll} 10^{-5}, & \mbox{r $<$ 2 {\rm
AU}} \\ 10^{-3}, & \mbox{2 {\rm AU} $<$ r $<$ 2.5 {\rm AU}} \\ 5\%, & \mbox{r
$>$ 2.5 {\rm AU}} \\ \end{array} \right.  
\end{equation}

This water distribution is imprinted on planetesimals and embryos at the start
of each simulation.  During accretion the water content of each body is
calculated by a simple mass balance of all the accreted bodies.  We do not take
into account water loss during giant impacts (Genda \& Abe 2005; Canup \&
Pierazzo 2006) or via hydrodynamic escape (Matsui \& Abe 1986; Kasting 1988). 

\subsection{Numerical Method}

Each simulation was integrated for at least 200 Myr using the hybrid
symplectic integrator {\tt Mercury} (Chambers 1999).  We used a 6-day timestep
for all integrations; numerical tests show that this is adequate to resolve
the innermost orbits in our simulations and to avoid any substantial error
buildup (see Rauch \& Holman 1999).  Collisions are treated as inelastic
mergers, and we assumed physical densities of $3 \, g \, cm^{-3}$ for all
embryos and planetesimals. Simulations were run on individual machines in a
distributed computing environment, and required 2-4 months per simulation.
The Sun's radius was artificially increased to 0.1 AU to avoid numerical error
for small-perihelion orbits.

For each Jupiter-Saturn-disk configuration we performed four different
simulations to account for the stochastic nature of accretion (e.g., Chambers
\& Wetherill 1998).  These four cases varied in terms of the random number used
to initialize our disk code, resulting in differences in the detailed initial
distributions of embryos and planetesimals.

Embryo particles interacted gravitationally with all other bodies but
planetesimal particles did not interact with each other.  This approximation
was made to reduce the run time needed per simulation which is already
considerable (see Raymond \etal 2006a for a discussion of this issue).  The
run time $\tau$ scales with the number of embryos $N_e$ and the number of
planetesimals, $N_p$, roughly as $\tau \sim N_e^2 + 2 N_e N_p$.  The
non-interaction of planetesimals eliminates an additional $N_p^2$ term.  Note
that $\tau$ refers to the computing time needed for a given timestep.  The
total runtime is $\tau$ integrated over all timesteps for all surviving
particles.  Thus, a key element in the actual runtime of a simulation is the
mean particle lifetime.  Configurations with strong external perturbations
(e.g., eccentric giant planets) tend to run faster because the mean particle
lifetime is usually shorter than for configurations with weak external
perturbations.

\section{Two Contrasting Examples}

\begin{figure*}
\centerline{\epsscale{1.}\plotone{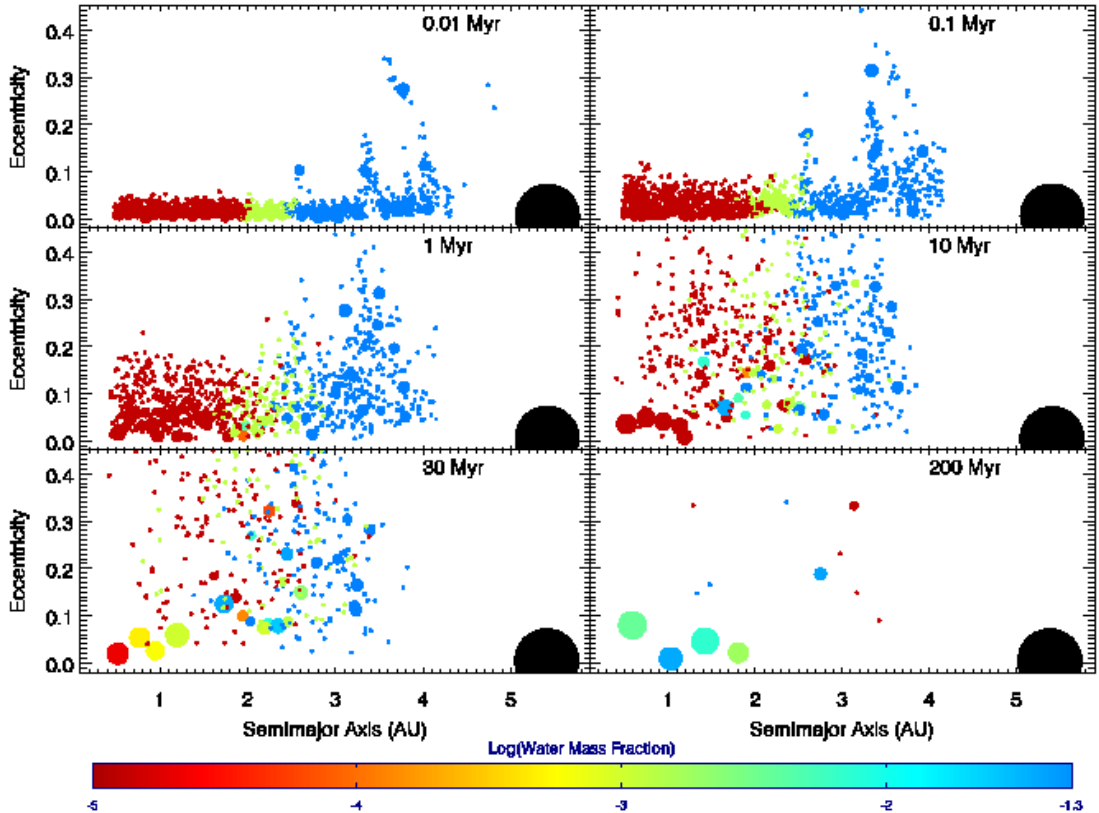}}
\caption{Snapshots in time from a simulation with Jupiter and Saturn in 3:2
mean motion resonance (JSRES).  The size of each body is proportional to its
mass$^{(1/3)}$ (but is not to scale on the x axis).  The color of each body
corresponds to its water content by mass, from red (dry) to blue (5\% water).
Jupiter is shown as the large black dot; Saturn is not shown. }
\label{fig:aet_jsres}
\end{figure*}

\begin{figure*}
\centerline{\epsscale{1.}\plotone{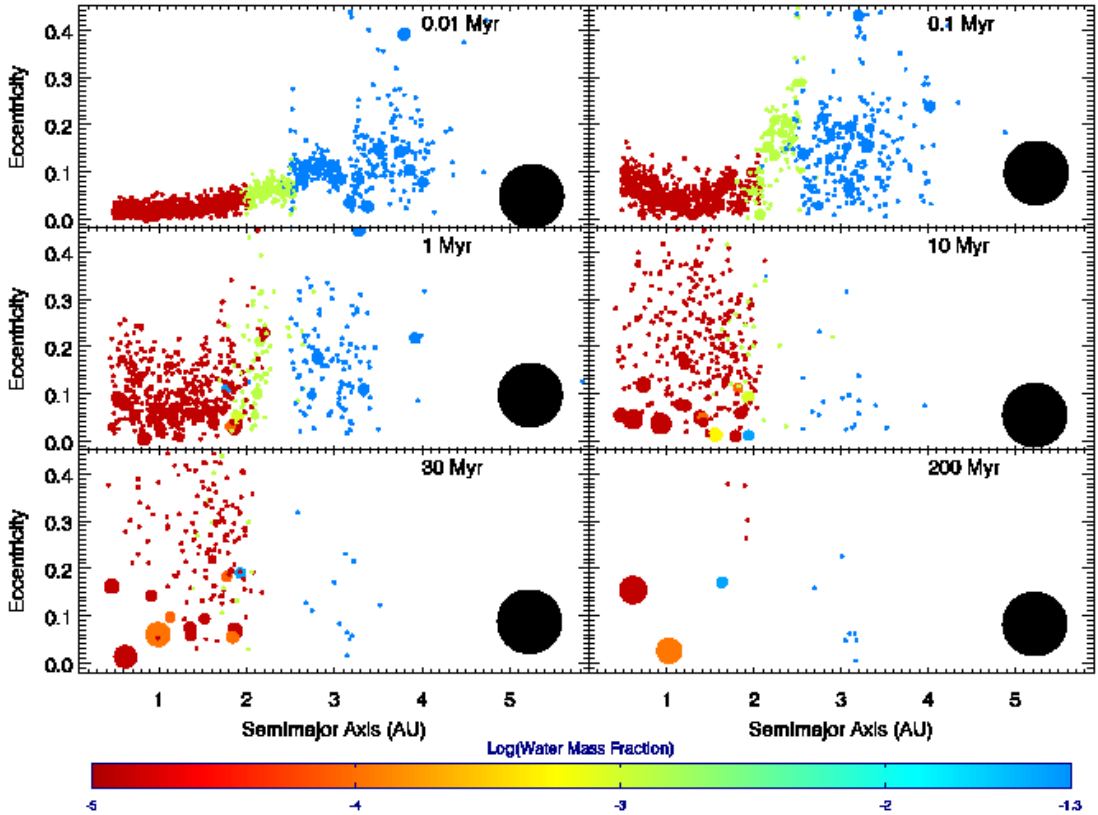}}
\caption{Evolution of a simulation with Jupiter and Saturn starting at their
current semimajor axes but with eccentricities of 0.1 (EEJS).   Formatted as
in Fig~\ref{fig:aet_jsres}.} 
\label{fig:aet_eejs}
\end{figure*}

\begin{deluxetable*}{ccccccc}
\tablewidth{0pt}
\tablecaption{Planets that formed in the JSRES and EEJS example simulations}
\renewcommand{\arraystretch}{.6}
\tablehead{
\colhead{Planet} &  
\colhead{a (AU)} &
\colhead{e\tablenotemark{1}} &
\colhead{i (deg)} &
\colhead{Mass ($\mearth$)} & 
\colhead{$WMF_\oplus$} &
\colhead{Last giant impact (Myr)}}
\startdata

JSRES-a &  0.59 & 0.08 & 1.7 & 0.95 & $2.77\times 10^{-3}$ & 113.5 \\
JSRES-b &  1.03 & 0.03 & 2.8 & 0.54 & $2.87\times 10^{-2}$ & 160.0\\
JSRES-c &  1.42 & 0.03 & 2.5 & 0.85 & $5.48\times 10^{-3}$ & 124.1\\
JSRES-d &  1.81 & 0.02 & 4.7 & 0.36 & $1.42\times 10^{-3}$ & 42.9\\
\\
EEJS-a & 0.61 & 0.08 & 3.3 & 0.90 &  $1\times10^{-5}$ &  82.2\\
EEJS-b & 1.02 & 0.05 & 3.2 & 0.70 &  $7.14\times 10^{-5}$ &  35.6\\
EEJS-c & 1.63 & 0.16 & 9.0 & 0.06 &  $3.08\times 10^{-2}$ & 0.168\\

\enddata
\tablenotetext{1}{Orbital values ($a$, $e$, $i$) are averaged over the last 1
Myr of each simulation.}
\end{deluxetable*}

We illustrate the variations between different cases using two simulations
with different configurations of Jupiter and Saturn: one case from the JSRES
batch and one from EEJS (simulations JSRES-4 and EEJS-3 in Table 2).  Each
simulation matched some of our constraints but neither matched all of them.
Figures~\ref{fig:aet_jsres} and~\ref{fig:aet_eejs} show snapshots in the
evolution of the two simulations.  Properties of the planets that formed in
each case are listed in Table 1.  We note that these are individual
simulations, and that there exists substantial variability in outcome between
simulations even for the same giant planet configuration.  We discuss the
outcomes of all simulations in section 5. 

In the JSRES simulation (Fig.~\ref{fig:aet_jsres}), eccentricities are excited
in the inner disk by interactions between embryos and planetesimals.  In the
outer disk, eccentricities are excited by specific mean motion resonances
(MMRs) with Jupiter and Saturn: the 3:1, 2:1 and 3:2 MMRs are clearly visible.
Eccentric embryos perturb nearby bodies and act to spread out the resonant
excitation on a Myr timescale.  A stage of chaotic growth lasts for $\sim$100
Myr.  During this time, there is substantial mixing of objects between radial
zones, the inner system is cleared of small bodies, and four water-rich
planets are formed inside 2 AU with masses between 0.36 and 0.95 $\mearth$
(see Table 1).

In the EEJS simulation (Fig.~\ref{fig:aet_eejs}), the inner and outer portions
of the disk are quickly divided by a strong secular resonance near 2 AU
($\nu_6$).  The evolution of the inner disk proceeds in similar fashion to the
JSRES simulation, although eccentricities are higher because of excitation by
another secular resonance at 0.7 AU ($\nu_5$).  The asteroid belt region was
cleared more quickly than for the JSRES case due to stronger secular and
resonant perturbations.  The stage of chaotic growth also lasts about 10$^8$
years but with less mixing between radial zones.  At the end of the
simulation, three mainly dry planets have formed within 2 AU.  The outermost
planet lies at 1.63 AU and is a good Mars analog.

\begin{figure*}
\centerline{\epsscale{0.5}\plotone{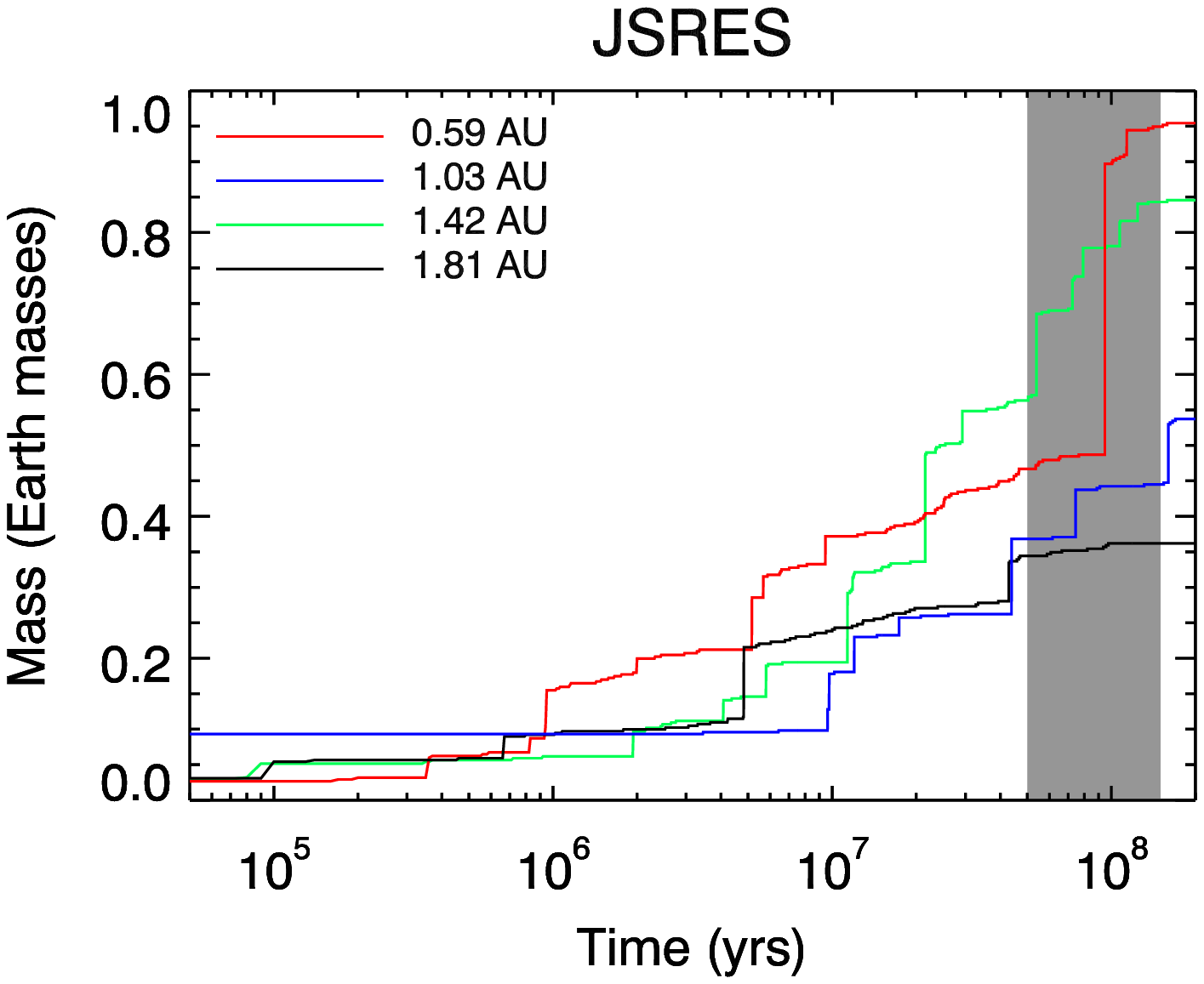}\plotone{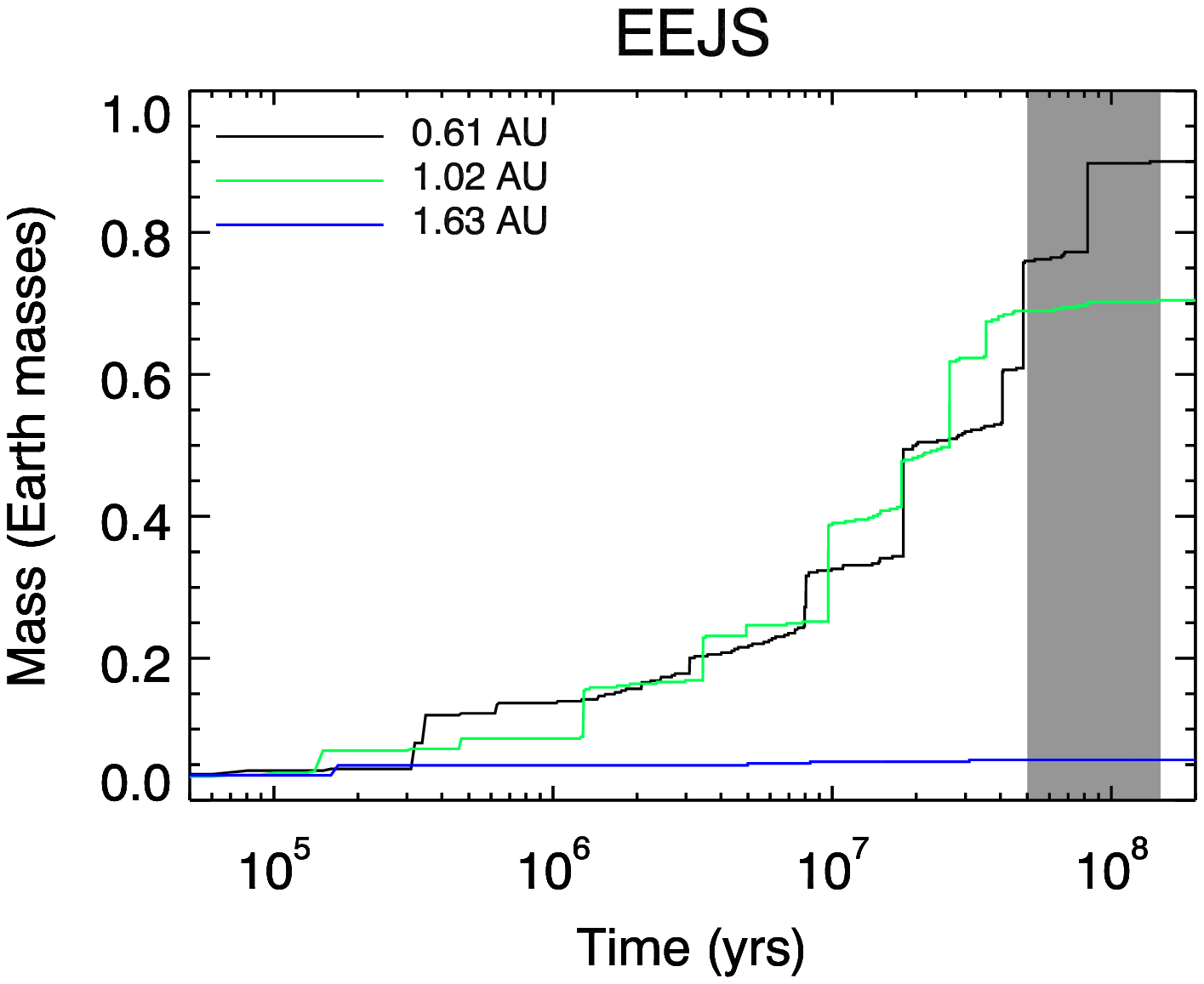}}
\centerline{\epsscale{0.5}\plotone{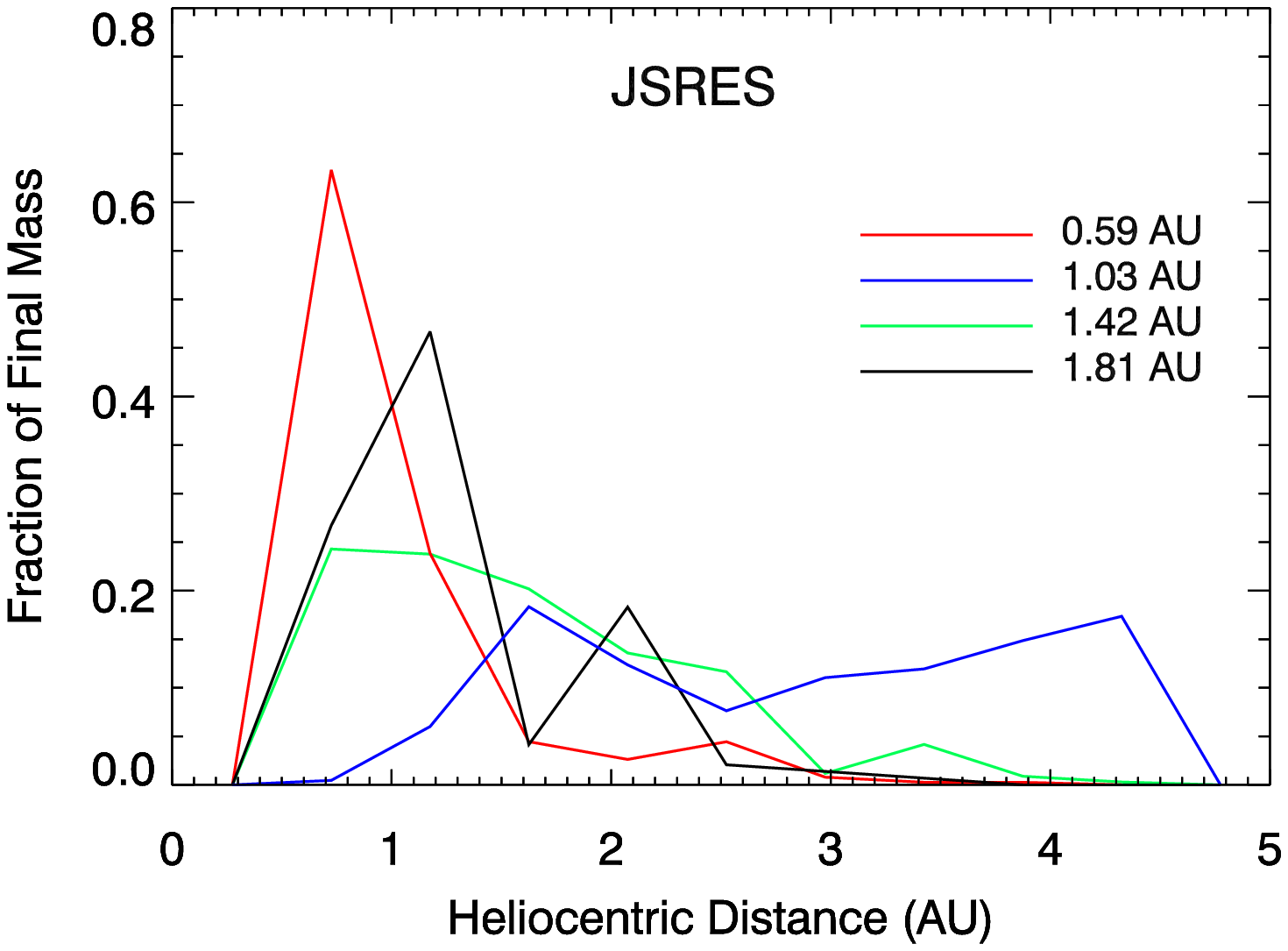}\plotone{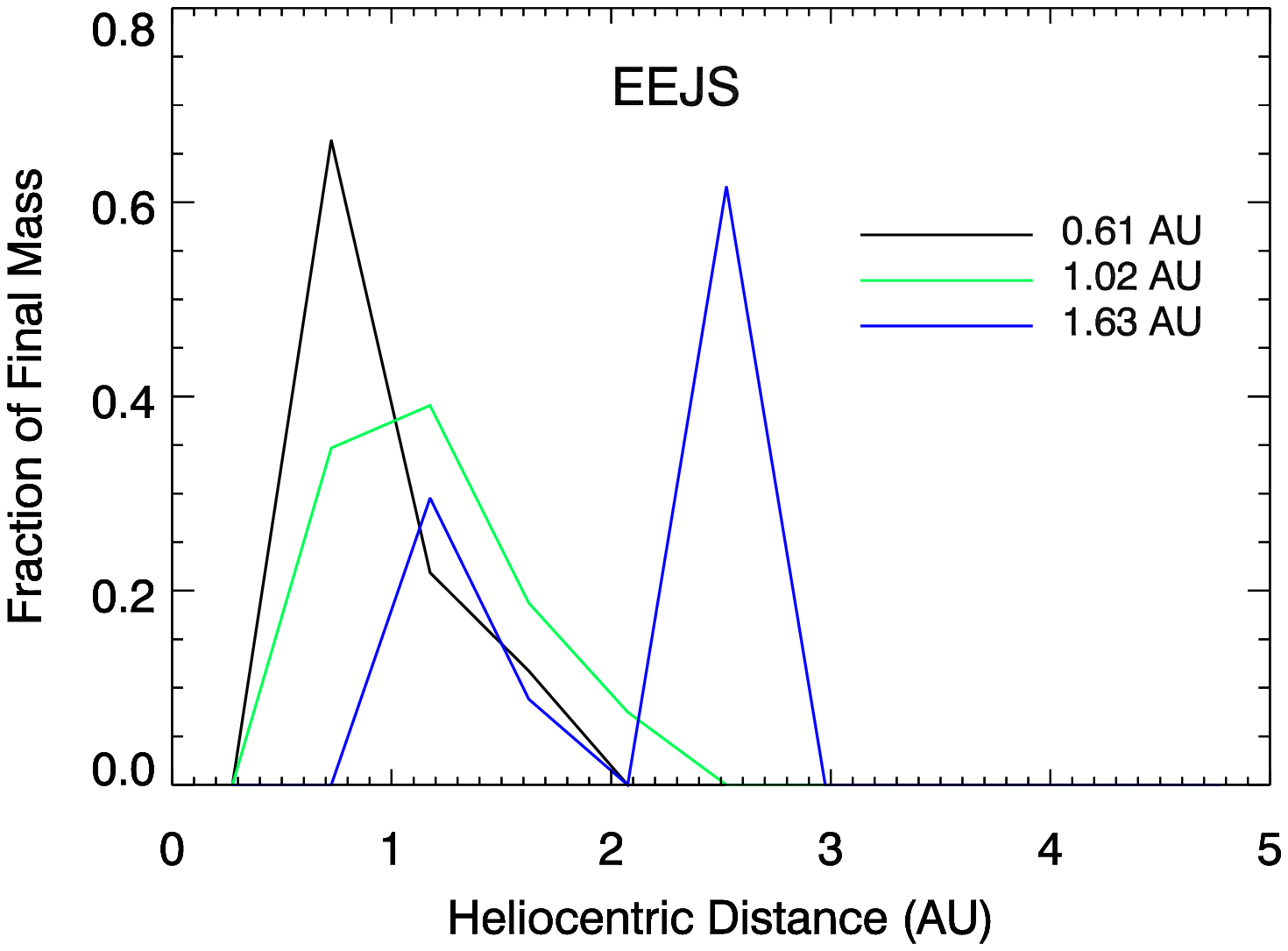}}
\caption{Growth and feeding zones for the planets that formed in our example
JSRES and EEJS simulations.  The top panels show the growth of each surviving
planet interior to 2 AU for the JSRES (left) and EEJS (simulations).  The
50-150 Myr isotopic constraints on the timing of the Moon-forming impact on
Earth are shaded.  Each colored curve corresponds to an individual planet, as
labeled.  The bottom panels show the origin of the material incorporated into
the planets.  See Table 1 and the text for details. }
\label{fig:fzone}
\end{figure*}

Figure~\ref{fig:fzone} (top panels) shows the mass of the planets over the 200
Myr span of the simulation for the two simulations.  Planetary growth is a
combination of relatively smooth accumulation from a large number of
planetesimals and punctuated accretion from a small number of giant impacts
with other embryos.  In general, embryo-embryo collisions increase in
magnitude in time simply because all embryos are growing.  This is
particularly clear for the case of the innermost planet (0.59 AU) in the JSRES
simulation which was hit by a 0.41 $\mearth$ embryo at 94.8 Myr while the
planet was only 0.48 $\mearth$.	

The timescale for the last giant impact on the JSRES planets was 43-160 Myr,
and 0.17-82 Myr for the EEJS planets.  The Earth analog (i.e., the planet
closest to 1 AU) in each simulation fell slightly out of the 50-150 Myr window
for the last giant impact on Earth (Touboul \etal 2007), but on different
sides.  The JSRES Earth analog's last giant impact was slightly too late (160
Myr) while the EEJS Earth analog's was too early (35.6 Myr).  The Mars analog
in the JSRES simulation (at 1.42 AU) has a mass that is roughly eight times
too large and a formation timescale that is far too long (124 Myr as compared
with the Hf/W isotopic age of 1-10 Myr; Nimmo \& Kleine 2007).  In contrast,
the EEJS simulation produced an excellent Mars analog that is actually
somewhat smaller than Mars (0.06 $\mearth$ vs. 0.11 $\mearth$) and whose only
giant impact occurred 168,000 years into the simulation.  Given that the
``time zero'' for our simulations is probably 1-3 Myr after the formation of
the Solar Nebula, this is consistent with isotopic measurements.  It is
interesting to note that the last giant impact on the innermost planet in each
simulation occurred quite late, at $\sim 10^8$ years (see Table 1).  The
reason for the late impact was different for the two simulations.  For the
JSRES simulation the last giant impactor originated in the asteroid belt,
where the timescale for close encounters and scattering is longer than the
inner system.  For the EEJS simulation, the last giant impactor originated at
1.2 AU but had its inclination increased by a short time spent in the vicinity
of the $\nu_6$ secular resonance, thereby prolonging its dynamical lifetime in
the inner system.  These late giant impacts on close-in planets contrast with
the nominal view of accretion occurring fastest in the inner regions of the
disk, especially given the much shorter accretion timescales for the Earth and
Mars analogs in the EEJS simulation.

The feeding zones of the planets from the JSRES and EEJS simulations are shown
in the bottom panels of Fig.~\ref{fig:fzone}.  Feeding zones were calculated
as the fraction of material incorporated into each planet that originated in
each 0.45 AU-wide radial bin.  The feeding zones of all planets overlap in
each simulation, although the width of individual feeding zones
vary.\footnote{Terrestrial feeding zones are not static, but actually widen
and move outward in time (Raymond \etal 2006a).}  In the JSRES simulation each
of the four planets accreted material from a radial width of more than 3 AU.
Given that the source of water lies beyond 2-2.5 AU, this explains the large
water abundance in the JSRES planets.  The Earth analog's feeding zone is
exceedingly wide and is unusual in that its accretion seed actually started
the simulation in the outer asteroid belt, at 4.3 AU.\footnote{A planet's
accretion seed is simply the object that was the larger in each of its
collisions.  The planet retains the name of this object.}  The Earth analog's
water content was therefore very large, roughly 30 times the Earth's current
water content without accounting for any water loss (the Earth's $WMF$ is
$\sim 10^{-3}$; Lecuyer \etal 1998). In contrast, the three planets from the
EEJS simulations each had feeding zones of less than 1.7 AU in width.  Very
little material from exterior to 2 AU was incorporated into the EEJS planets,
with the notable exception of one embryo that originated at 2.64 AU and was
the accretion seed of the Mars analog. Thus, the Earth and Venus analogs are
very dry, but the Mars analog is very water-rich.

The JSRES and EEJS simulations each reproduced some of our constraints but
neither reproduced them all.  The JSRES simulation formed a terrestrial planet
system with eccentricities and inclinations almost as low as the Solar
System's terrestrial planets', with an $AMD$ of 0.0023 (as compared with
0.0018 for Mercury, Venus, Earth and Mars -- hereafter MVEM).  The JSRES
planets also contain abundant water that was delivered from the primordial
asteroid belt.  The formation timescale of the Earth analog is roughly
consistent with isotopic constraints for Earth. However, the JSRES Mars analog
bears little resemblance to the real planet in terms of its mass and formation
timescale.  In addition, three extra large bodies exist at the end of the
simulation: a 0.36 $\mearth$ planet at 1.8 AU and two embryos in the asteroid
belt totaling 0.11 $\mearth$.  These remnant bodies, in particular the embryos
in the asteroid belt, are inconsistent with the observed inner Solar System.

The EEJS planets are novel among accretion simulations of this kind because
they contain a reasonable Mars analog in terms of its mass, orbit and
formation timescale.  In addition, the approximate masses and spacing of the
EEJS planets are close to those of Venus, Earth and Mars.  No embryos are
stranded in the asteroid belt although a dozen planetesimals remain in the
belt.  However, the EEJS Earth analog's formation timescale is too short by
$\sim$ 20\%.  More importantly, the $AMD$ for the system is 0.0086, roughly 5
times higher than the MVEM value.  Finally, the Earth analog is almost
completely devoid of asteroidal water and thus requires an alternate source.

The values of the radial mass concentration statistic $RMC$ of both
simulations are far lower than for the inner Solar System.  The $RMC$ values
are 28.5 for the JSRES simulation and 44.2 for the EEJS case, as compared with
89.9 for MVEM.

\begin{figure}
\centerline{\epsscale{1.}\plotone{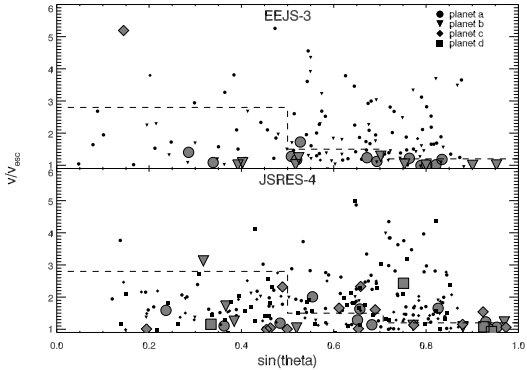}}
\caption{Impact angles and velocities for all collisions that occurred in our
example EEJS and JSRES simulations.  Large grey symbols refer to embryo-embryo
impacts and small black symbols to embryo-planetesimal impacts.  Each symbol
refers to the impacts that occurred on a specific planet -- for each
simulation, planets are ordered by their proximity to the star: planet a is
closest, followed by planet b, etc (the EEJS simulation only formed three
planets so there is no planet d).  Note that theta=0 and 90$^\circ$
corresponds to head-on and graxing impacts, respectively.  Impact velocities
are normalized by the two-body escape speed $v_{esc}$ (see Eqn 5).  The dashed
line is the approximate boundary between accretionary (below the line) and
erosive collisions (Agnor \& Asphaug 2004).}
\label{fig:coll}
\end{figure}

Figure~\ref{fig:coll} shows the details of each planetesimal and embryo
collision that occurred on the surviving planets in the two simulations.  The
impact angle theta is defined to be zero for a head-on collision and
90$^\circ$ for a grazing collision.  The impact velocity is given in terms of
the two-body escape speed $v_{esc}$ : \begin{equation} v_{esc} = \sqrt{\frac{2
G \left(M_1 + M_2\right)}{R_1+R_2}}, \end{equation} \noindent where $G$ is the
gravitational constant, $M_1$ and $M_2$ are the colliding bodies' masses, and
$R_1$ and $R_2$ are the bodies' physical radii. In the absence of 3-body
effects, which are relevant in at most a few percent of collisions, collisions
can only occur at $v/v_{esc} > 1$.  

Planetesimal-embryo impacts (small symbols in Fig.~\ref{fig:coll}) tend to
occur at higher velocities than embryo-embryo impacts simply because their
eccentricities are higher on average due to viscous stirring.  However, we
expect virtually all planetesimal-embryo collisions to result in net growth.
Note that our numerical scheme does not allow for planetesimal-planetesimal
collisions (see discussion in \S 6).

High-speed or off-center embryo-embryo collisions (large symbols in
Fig.~\ref{fig:coll}) can result in either partial accretion or even erosion.
Agnor \& Asphaug (2004) showed that accretionary collisions only occur at
$v/v_{esc} \lesssim 1.5$ and preferentially for small impact angles.  The
majority of giant impacts (filled circles in Fig.~\ref{fig:coll}) occur at low
speeds and should therefore be accretionary.  Indeed, the majority of impacts
lie in the accretionary regime as defined by Agnor \& Asphaug (2004): 60\%,
71\%, 71\%, and 75\% for the four planets from the JSRES simulation (listed
from closest- to farthest from the Sun), and 83\%, 100\% and 0\% for the three
EEJS planets.\footnote{Note that for the outermost EEJS planet (the Mars
analog), only one giant collision occurred, when the planet was just 0.036
$\mearth$ and was hit by a 0.015 $\mearth$ embryo.  The collision was nearly
head-on (sin{theta} = 0.14) but high-speed ($v/v_{esc}$ = 5.2).}  This is a
larger fraction than the 55\% found by Agnor \etal (1999) and the $\sim$ half
inferred by Agnor \& Asphaug (2004).  We assume that dynamical friction from
small bodies reduced the mean impact speed and increased the fraction of
accretionary impacts.  We note, however, that the definition of an
accretionary impact only requires that the collision produce an object larger
than either of the two impactors, not that the object's mass equal the sum of
the colliding masses.  In particular, each of our example simulations has a
cluster of impacts at low velocity but large angle (see discussion in section
5).  A large number of fragments is probably produced in these off-center
collisions (Asphaug \etal 2006).  We do not have the ability to track the
effect of these fragments, which could be important (see discussion in Section
6).  On the other hand, an erosive, ``hit and run'' collision usually results
in an extra body that looks very similar to the original impactor and can
easily be accreted in a later collision involving that extra body (Asphaug
\etal 2006).

\begin{figure}
\plotone{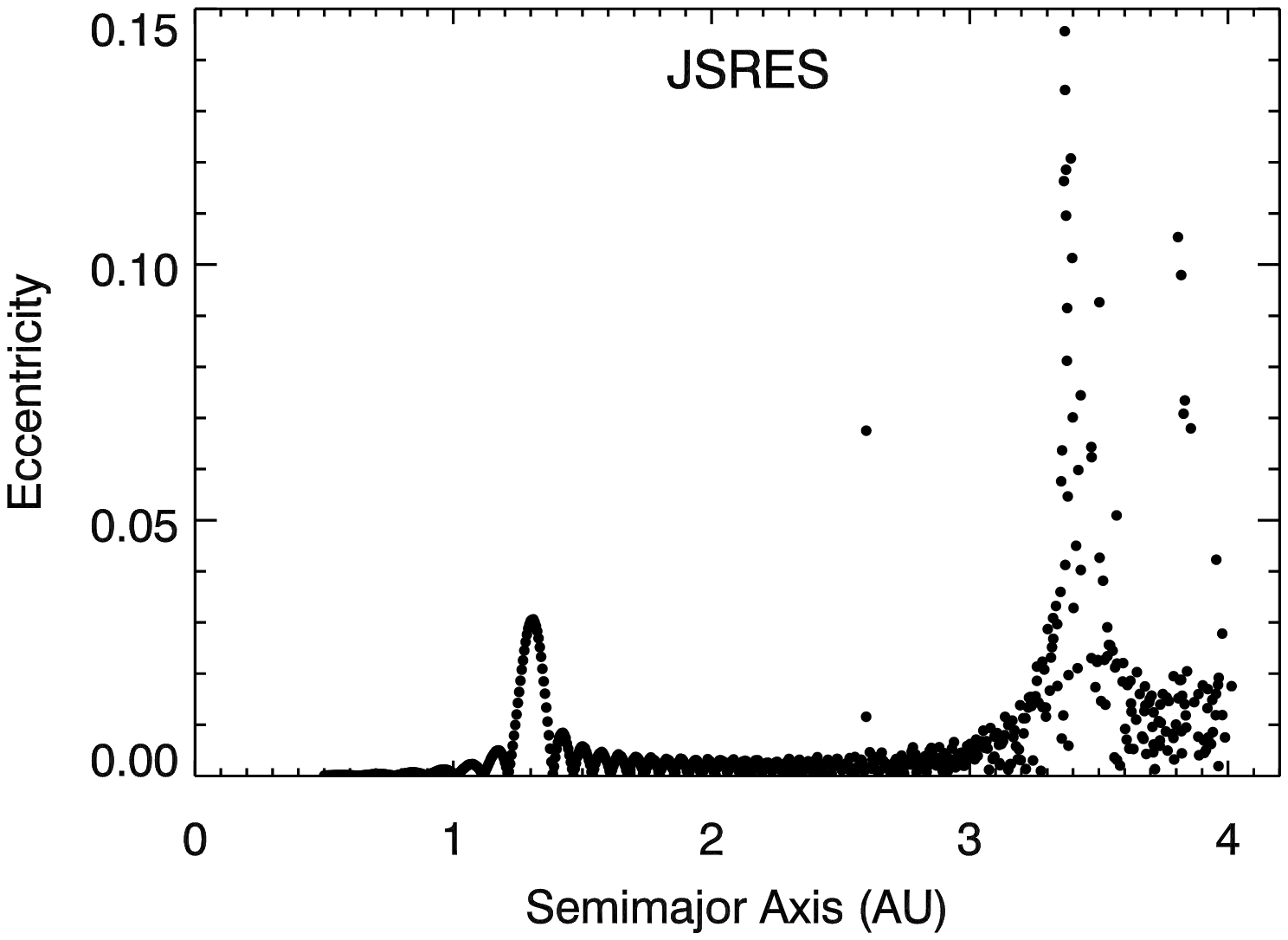}
\plotone{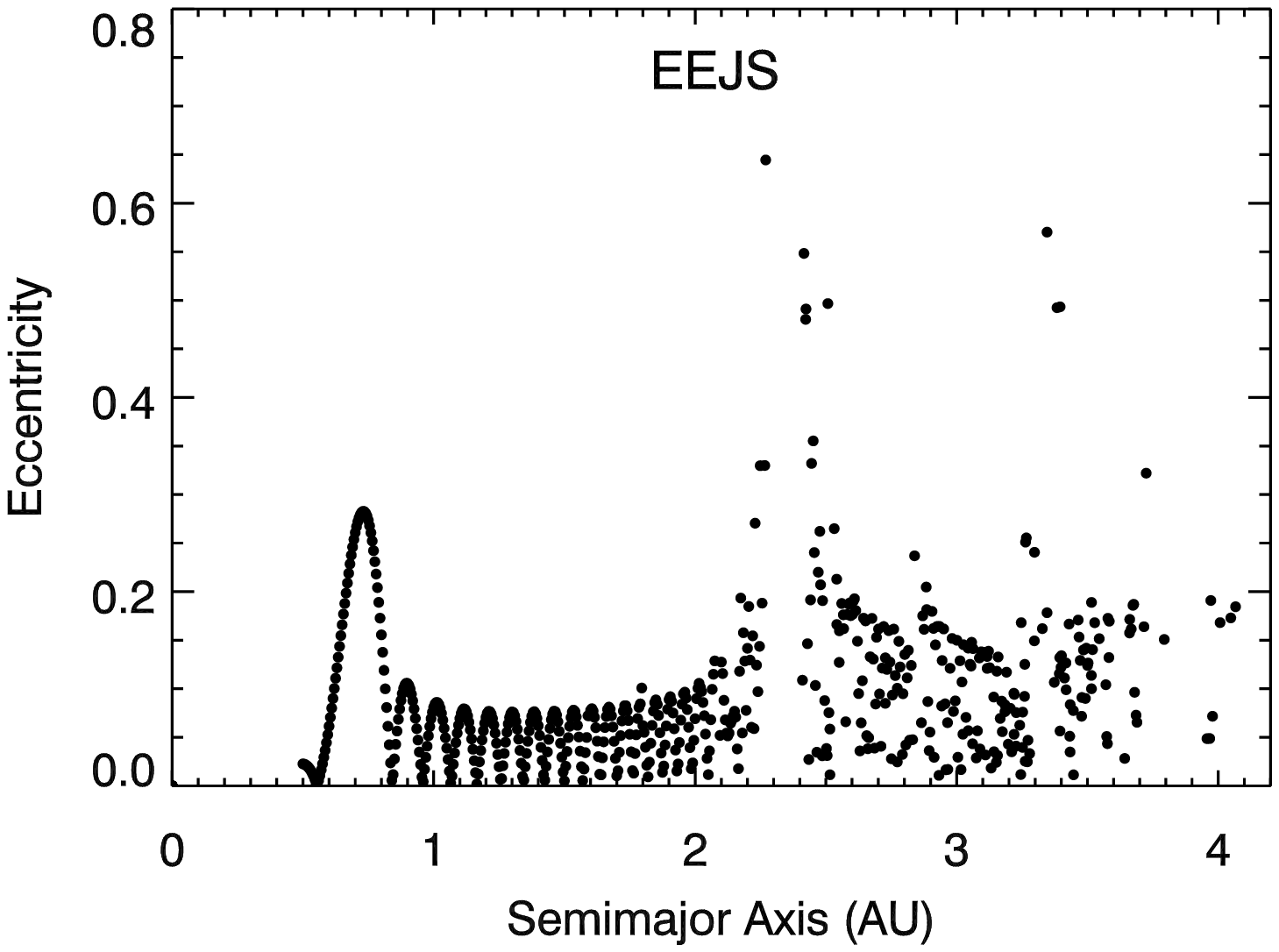}
\caption{Orbital eccentricities of massless test particles in
the inner Solar System after 1 Myr for the JSRES (left panel) and EEJS (right
panel) configurations of Jupiter and Saturn.  Note the dif\mbox{}ference in
the $y$ axis scale between the two panels.}
\label{fig:ae_tp}
\end{figure}

The different outcomes in the EEJS and JSRES example simulations can be
attributed to differences in eccentricity and inclination excitation by
specific resonances with Jupiter and Saturn, as well as by secular
perturbations from Jupiter and Saturn.  Figure~\ref{fig:ae_tp} shows the
eccentricities of test particles on initially circular orbits after 1 Myr of
evolution in each giant planet system, with no embryos present.  For the JSRES
case, the amount of eccentricity excitation is small.  The main sources of
excitation are the $\nu_5$ secular resonance at 1.3 AU and the 2:1 MMR with
Jupiter at 3.4 AU.  The 3:1 and 3:2 MMRs with Jupiter are faintly visible at
2.6 AU and 4.1 AU.  The small amount of external forcing means that the
self-scattering of embryos and planetesimals is the dominant source of
eccentricity in the JSRES simulations.  Given that the disk is continuous and
contains a significant amount of mass in the Mars region, no dynamical
mechanism exists to remove that mass.  In addition, the weak influence of the
giant planets allows for efficient delivery of water-rich material via a large
number of relatively weak embryo-embryo and embryo-planetesimal scattering
events (Raymond \etal 2007).  

In contrast, perturbations from Jupiter and Saturn play a dominant role in the
EEJS configuration.  Strong secular resonances are visible at 0.7 AU ($\nu_5$)
and 2.2 AU ($\nu_6$).  In addition, secular excitation is strong enough to
impart a typical free eccentricity of 0.1-0.2 throughout the inner Solar
System.  The $\nu_6$ secular resonance is directly responsible for Mars' small
size, as it efficiently removes mass from the 1.5-2.5 AU region, mainly by
driving eccentricities of bodies to 1 and inducing collisions with the Sun.
However, the $\nu_6$ acts as a barrier between the terrestrial planets and the
asteroid belt, such that water delivery is severely reduced.  The strong
eccentricity forcing throughout the inner Solar System appears to prevent
low-$AMD$ terrestrial planets from forming.  However, the scattering of
embryos and planetesimals by Jupiter and Saturn throughout accretion reduces
the giant planets' eccentricities and weakens their secular perturbations in
time.  

Despite the differences between the JSRES and EEJS simulations, it is
important to realize that small secular perturbations from the giant planets
do not necessarily correlate with low-AMD terrestrial planets, especially in
the case of limited numerical resolution.  In fact, O'Brien \etal (2006)
formed significantly lower-$AMD$ terrestrial planets for the EJS configuration
than their CJS simulations.  The reason for this is that the timescale for the
removal of asteroidal material was very long for the CJS simulations, and
close encounters with late-arriving material from the asteroid belt tended to
increase eccentricities.  In contrast, their EJS simulations cleared out the
asteroid belt quickly and the secular forcing of eccentricities was small
enough inside $\sim$ 2 AU that the giant planets did not act to increase the
terrestrial planets' $AMD$.

\section{Simulation Outcomes and Comparison with our Constraints}

The evolution of each simulation proceeded in a qualitatively similar fashion
to the example EEJS or JSRES simulations.  In fact, the two cases illustrated
in section 4 comprise the most extreme variations in our sample.  The other
cases lie between those extremes, typically with a moderate amount of
excitation from the giant planets in the outer disk and relatively little
external excitation in the inner disk.  In this section we discuss the
outcomes of our simulations in terms of how they compare with our Solar System
constraints.  We explore how the differences between cases can be attributed
to the giant planet configuration and, to a lesser degree, to variations in
the disk's density profile.

There was a large range in the characteristics of the terrestrial planet
systems that formed. The number of planets in a given system ranged from 2-6,
where we define a planet to contain an least one embryo, to be interior to 2
AU, and to be on a stable orbit that does not cross the orbit of any other
planets or embryos.  The total mass in planets varied by almost a factor of
two, from 1.4 to 2.7 $\mearth$.  Table 2 summarizes the outcome of each
simulation.

\scriptsize
\begin{deluxetable*}{l|ccccccccc}
\tablewidth{0pt}
\tablecaption{Comparison between simulations and observed constraints\tablenotemark{1}}
\renewcommand{\arraystretch}{.6}
\tablehead{
\colhead{Simulation} &  
\colhead{$N_p$} &
\colhead{$M_{tot}$} &
\colhead{$AMD$} &
\colhead{$RMC$} &
\colhead{$WMF_\oplus$} &
\colhead{$M_{Mars} (\mearth)$} &
\colhead{$Tf_\oplus$ (Myr)} &
\colhead{$Tf_M$} & 
\colhead{N(ast.emb.)}}
\startdata

CJS15-1       &  3 & 2.70 &  0.0027 &35.5 & $1.8\times 10^{-3}$ & 1.45 &    50.7 &   113.7 &  2 \\
CJS15-2       &  3 & 2.83 &  0.0107 &27.3 & $5.7\times 10^{-3}$ & 0.97 &   141.9 &    81.5 &  1 \\
CJS15-3       &  4 & 2.89 &  0.0030 &27.2 & $6.1\times 10^{-3}$ & 0.98 &    75.0 &   113.6 &  0 \\
CJS15-4       &  4 & 2.68 &  0.0030 &29.8 & $5.3\times 10^{-3}$ & 0.75 &   104.1 &    36.1 &  3 \\
CJS1-1 	      &  2 & 2.30 &  0.0166 &21.6 & $1.5\times 10^{-3}$ & 1.05 &   149.7 &   186.3 &  2 \\
CJS1-2 	      &  3 & 2.00 &  0.0315 &44.8 & $7.9\times 10^{-3}$ & 0.67 &   139.6 &   162.2 &  2 \\
CJS1-3 	      &  4 & 2.45 &  0.0019 &30.2 & $3.2\times 10^{-3}$ & 0.89 &    33.3 &   100.3 &  0 \\
CJS1-4 	      &  2 & 2.53 &  0.0104 &27.8 & $2.1\times 10^{-3}$ & 1.32 &   123.5 &   101.1 &  1 \\
CJSECC15-1    &  3 & 2.20 &  0.0047 &45.4 & $3.1\times 10^{-3}$ & 0.58 &    80.4 &    63.8 &  1 \\
CJSECC15-2    &  4 & 2.37 &  0.0053 &34.9 & $1.2\times 10^{-3}$ & 0.59 &    75.5 &    46.1 &  3 \\
CJSECC15-3    &  4 & 2.42 &  0.0030 &37.5 & $6.9\times 10^{-4}$ & 1.09 &    96.3 &   164.1 &  5 \\
CJSECC15-4    &  3 & 2.27 &  0.0010 &40.9 & $9.3\times 10^{-4}$ & 0.69 &    29.2 &    78.1 &  2 \\
EJS15-1       &  3 & 2.08 &  0.0018 &34.9 & $1.7\times 10^{-4}$ & 0.81 &    56.1 &    76.3 &  2 \\
EJS15-2       &  2 & 2.03 &  0.0025 &48.9 & $3.3\times 10^{-4}$ & -- &   812.3 &   -- &  1 \\
EJS15-3       &  3 & 2.05 &  0.0050 &44.2 & $1.9\times 10^{-4}$ & 0.26 &    38.9 &   118.3 &  1 \\
EJS15-4       &  4 & 2.07 &  0.0062 &34.7 & $2.6\times 10^{-4}$ & 0.11 &    65.7 &    41.9 &  1 \\
EJS1-1 	      &  2 & 1.66 &  0.0063 &39.5 & $1.5\times 10^{-4}$ & -- &   147.8 &   -- &  1 \\
EJS1-2 	      &  3 & 1.43 &  0.0101 &46.0 & $6.3\times 10^{-3}$ & 0.43 &   565.6 &   190.6 &  1 \\
EJS1-3 	      &  3 & 1.60 &  0.0124 &40.5 & $7.7\times 10^{-4}$ & 0.22 &   142.0 &   548.6 &  1 \\
EJS1-4 	      &  2 & 1.51 &  0.0035 &51.2 & $1.4\times 10^{-2}$ & -- &   169.2 &   -- &  1 \\
EEJS-1      &  3 & 1.83 &  0.0178 &33.6 & $1.0\times 10^{-5}$ & 0.34 &   109.3 &    59.6 &  1 \\
EEJS-2      &  3 & 1.67 &  0.0151 &50.9 & $1.1\times 10^{-4}$ & 0.16 &    59.5 &     8.1 &  1 \\
EEJS-3      &  3 & 1.66 &  0.0086 &63.9 & $8.1\times 10^{-5}$ & 0.06 &    35.6 &     0.2 &  0 \\
EEJS-4      &  3 & 1.89 &  0.0112 &43.6 & $2.1\times 10^{-5}$ & 0.07 &   165.0 &   0.2 &  0 \\
EEJS-5      &  4 & 1.78 &  0.0279 &39.5 & $1.5\times 10^{-5}$ & 0.34 &   102.3 &   128.1 &  0 \\
EEJS-6      &  5 & 1.87 &  0.0099 &39.4 & $2.9\times 10^{-5}$ & 0.40 &   116.7 &    26.4 &  1 \\
EEJS-7      &  3 & 1.85 &  0.0038 &33.9 & $3.6\times 10^{-3}$ & 0.44 &   129.4 &    98.3 &  2 \\
EEJS-8      &  3 & 1.91 &  0.0248 &31.6 & $8.6\times 10^{-3}$ & 0.32 &   199.8 &     9.1 &  0 \\
EEJS-9      &  5 & 1.83 &  0.0027 &42.1 & $2.3\times 10^{-3}$ & 0.23 &    33.1 &    91.3 &  1 \\
EEJS-10     &  3 & 1.84 &  0.0047 &42.6 & $1.9\times 10^{-5}$ & 0.58 &    55.1 &   166.3 &  1 \\
EEJS-11     &  4 & 1.72 &  0.0033 &49.0 & $1.2\times 10^{-4}$ & 0.09 &   145.0 &     4.5 &  1 \\
EEJS-12     &  4 & 1.81 &  0.0027 &44.9 & $2.5\times 10^{-4}$ & 0.20 &    45.2 &     2.3 &  0 \\
JSRES-1       &  6 & 2.61 &  0.0022 &32.4 & $7.5\times 10^{-3}$ & 0.54 &    29.2 &    28.1 &  2 \\
JSRES-2       &  2 & 2.31 &  0.0119 &61.0 & $5.3\times 10^{-3}$ & 1.27 &   176.8 &   176.8 &  6 \\
JSRES-3       &  4 & 2.55 &  0.0071 &34.2 & $1.2\times 10^{-3}$ & -- &   115.3 &    -- &  1 \\
JSRES-4       &  4 & 2.70 &  0.0023 &28.5 & $2.9\times 10^{-2}$ & 0.85 &   160.1 &   124.1 &  2 \\
JSRESECC-1    &  4 & 2.70 &  0.0016 &28.7 & $1.3\times 10^{-3}$ & 1.01 &    20.2 &    81.0 &  1 \\
JSRESECC-2    &  4 & 2.73 &  0.0044 &26.8 & $3.5\times 10^{-4}$ & 0.96 &    73.7 &    60.6 &  1 \\
JSRESECC-3    &  4 & 2.51 &  0.0041 &28.1 & $3.0\times 10^{-4}$ & 0.73 &    78.5 &    99.8 &  3 \\
JSRESECC-4    &  3 & 2.48 &  0.0025 &39.3 & $1.1\times 10^{-3}$ & 0.98 &   110.3 &   177.9 &  2 \\
\hline
\\
MVEM\tablenotemark{3} & 4 & 1.98 & 0.0018 & 89.9 & $\sim 1\times 10^{-3}$ & 0.11 & 50-150 & 1-10& 0
\\
\enddata
\tablenotetext{1}{Table columns are: the simulation, the number of terrestrial
planets inside 2 AU $N_{pl}$, the total mass in those planets $M_{tot}$, the
angular momentum deficit $AMD$ (see Eqn 2), the radial mass concentration
statistic $RMC$ (see Eqn 1), the water content by mass of the simulation's
Earth analog $WMF_\oplus$, the mass of the simulation's Mars analog, the
time of the last giant impact on the Earth and Mars analogs $Tf_\oplus$ and
$Tf_M$, and the number of embryos stranded in the asteroid belt that were more
massive than 0.03 $\mearth$.  A comparison with the Solar System's terrestrial
planets (MVEM) is shown at the bottom.}
\tablenotetext{2}{A few simulations did not form Mars analogs at all (i.e., no planets between 1.25 and 1.75 AU). }
\tablenotetext{3}{Earth's water content is not well known because the amount of
water in the mantle has been estimated to be between 1-10 ``oceans'', where 1
ocean ($= 1.5\times 10^{24} g$) is the amount of water on Earth's surface
(Lecuyer \etal 1998).  Our estimate of $10^{-3}$ for Earth's water content by
mass assumes that 3 oceans are locked in the mantle.}
\end{deluxetable*}
\normalsize

\begin{figure}
\centerline{\epsscale{1.}\plotone{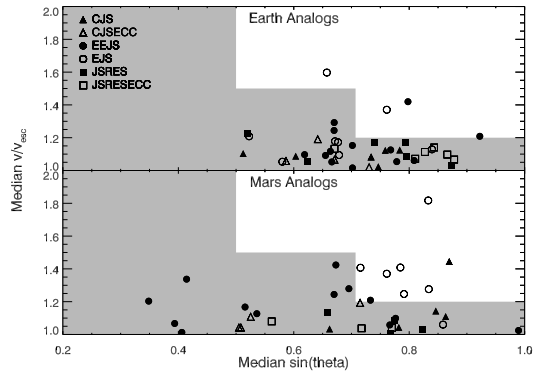}}
\caption{Median angles and velocities for the giant collisions that formed the
Earth and Mars analogs in each simulation, labeled with different symbols (see
legend).  The shaded region represents the zone where impacts should be
accretionary rather than erosive (Agnor \& Asphaug 2004).  Recall that theta=0
and 90$^\circ$ corresponds to head-on and graxing impacts, respectively.
Impact velocities are normalized by the two-body escape speed $v_{esc}$ (see
Eqn 5). }
\label{fig:coll_all}
\end{figure}

Figure~\ref{fig:coll_all} shows the median collision velocities and angles for
the giant (embryo-embryo) collisions that occurred during the formation of the
Earth and Mars analogs in each of our simulations.\footnote{Earth and Mars
analogs are defined to be the most massive planets in the region from 0.8-1.25
AU, and 1.25-1.75 AU, respectively.  If no planet exists in that zone, then
the Earth analog is taken to be the planet that is closest to 1 AU and the
Mars analog is taken to be the outermost planet inside 2 AU.}  Despite the
existence of high-velocity impacts (see Fig.~\ref{fig:coll}), the median
collisional values are quite modest and in almost all cases the vast majority
of giant collisions are accretionary rather than erosive.

If typical impact speeds or angles on Mars analogs were much higher than
for Earth analogs, then the fraction of erosive collisions on Mars analogs
would be higher and one could claim that the mass ratio of Mars- to
Earth-analogs in our simulations was too high.  However, the distribution
of collision velocities for Earth and Mars analogs is very similar
(Fig.~\ref{fig:coll_all}).  Therefore, we can rule out variations in impact
properties as the source of large Mars analogs in our simulations.  One
exception are the EJS simulations, many of which have somewhat higher
impact speeds and angles for Mars analogs than for Earth analogs.  For the
EJS simulations, the mass of Mars analogs may therefore be somewhat
overestimated.

Canup (2004) showed that a very particular impact configuration was required
to form the Moon.  Such an impact must be low-velocity ($v/v_{esc} < 1.1$),
off-center (sin[theta] between 0.67 and 0.76), and have an impactor to target
mass ratio between 0.11 and 0.15.  Canup (2008) found that, for prograde
rotation of the proto-Earth, slightly smaller impactors can form Moon-analogs,
with a cutoff at roughly 0.1.  For retrograde rotation of the proto-Earth,
larger impactor-to-target mass ratios are allowed but ratios less than 0.1 are
still unable to form the Moon.  We examined the last three giant impacts
suffered by the Earth analog in each simulation, and none of the impacts
fulfilled Canup's (2004) three requirements.  In fact, none of the last three
impacts on an Earth analog had an impactor to target mass ratio larger than
0.05 for the right collision angle and speed.  In addition, only 4\% of the
late giant impacts satisfied Canup's velocity and angle criteria.  We conclude
that the Earth's Moon must be a cosmic rarity unless differences between
planetary systems produce a systematic change in the likelihood of
Moon-forming impacts.  However, we note that the simulations of Canup (2004,
2008) were specifically designed to reproduce the details of the Earth-Moon
system, in particular its high specific angular momentum and the Moon's small
core.  We suspect that a much larger range of late giant collisions would
produce satellites, although their properties could be much different than the
Moon.

As noted in section 4, our simulations show a grouping of low-velocity grazing
collisions.  The difference between the velocity distributions at small and
large angles is only significant for grazing collisions with sin (theta) $>$
0.9, where the collisions that were registered by the code do indeed occur at
lower speeds.  When comparing the statistics of head-on (sin [theta] $<$ 0.7)
and grazing (sin [theta] $>$ 0.9) collisions in all 40 simulations, there were
no notable differences in terms of collision time, distance from the Sun, or
the details of the impactor.  Grazing collisions did, however, have target
masses $\sim$ 15\% larger than for head-on collisions.  Statistically, one
would expect collisions between equal-mass objects to have a wider
distribution in sin(theta) than for collisions with one dominant mass,
although this also depends on the collision speed.  The small mass ratio for
the grazing collisions may explain the low collision speeds, simply because
the two-body escape speed is larger than for unequal-mass objects.

It is possible that the {\tt Mercury} code (Chambers 1999) has difficulty
registering high-speed grazing collisions because they could travel many Hill
radii in a single timestep.  For example, two bodies traveling with relative
velocity of 10 $km \, s^{-1}$ travel 0.035 AU with respect to each other in a
single 6 day timestep.  The Earth's Hill sphere $R_H$ is about 0.01 AU, and
approaches within $3 R_H$ are tracked numerically with the Bulirsch-Stoer
method rather than the symplectic map.  Thus, any two objects that are flagged
as having a close encounter will have their orbits faithfully resolved.
However, if the two objects were never flagged to approach within $3 R_H$ then
an encounter could be missed.  If that were the case then we would expect to
miss more collisions at small orbital distances because the Hill sphere is
smaller and relative velocities are larger.  Although our statistics are
limited, we don't see any evidence for this.  On the other hand, the easiest
grazing collisions for {\tt Mercury} to find should be those between massive
bodies traveling at low speeds, and we can think of no obvious physical reason
that high-speed grazing collisions should not occur.  Thus, although we have
not found any evidence of the code missing high-speed grazing collisions, we
can not rule out the possibility.  We expect that such collisions would likely
result in a ``bounce'' rather than a collision (Asphaug \etal 2006), and that
later lower-speed or head-on collisions, impacts that are certainly found by
{\tt Mercury}, would be the ones to result in planetary growth.

The effect of varying the disk's surface density profile between $r^{-1}$
(i.e., $x=1$ from Eqn. 3) and $r^{-1.5}$ ($x=3/2$) not insignificant, although
we only varied this parameter for the CJS and EJS configurations of Jupiter
and Saturn (EJS1 and CJS1 had $r^{-1}$, EJS15 and CJS15 had $r^{-1.5}$).  The
properties of the CJS1, CJS15, EJS1 and EJS15 simulations are summarized in
Table 3.  The $r^{-1}$ simulations formed slightly fewer planets, contained
less total mass in planets, had longer formation timescales for Earth and
higher $AMD$ values for the final systems than the $r^{-3/2}$ simulations.  In
addition, Earth analogs in the EJS1 simulations contained far more water than
Earths in the EJS15 simulations.\footnote{The same trend was not seen in the
CJS1 vs. CJS15 simulations because the small number of planets that formed in
the CJS1 cases led to the Earth analog being located at $\sim$ 0.8 AU in 3/4
cases.  Given that water delivery decreases with distance from the water
source ($>2-2.5$ AU), this decreased the water content of Earth analogs in the
CJS1 simulations.}  These trends are consistent with the results of Raymond
\etal (2005), and appear to be due simply to the fact that the $r^{-1}$
simulations contain far more mass in the asteroid region than the $r^{-1.5}$
simulations.  Given that planets form largely from local material, the
$r^{-1}$ simulations contain less material in the inner disk and therefore
form less massive planets.  In addition, the relatively large amount of
water-rich material in the asteroid belt increases the probability of water
delivery, although water delivery is more sensitive to the giant planet
configuration than to the disk properties: despite the large median value,
note that 2/4 EJS1 simulations formed Earths with less than 1 part per
thousand of water (see Table 2).  The large amount of asteroidal material in
the $r^{-1}$ simulations also prolongs the period of chaotic bombardment,
increasing the mean formation time for Earth.

\scriptsize
\begin{deluxetable*}{l|ccccc}[t]
\tablewidth{0pt}
\tablecaption{Mean properties of terresrtial planet systems for different disk
surface density profiles\tablenotemark{1}}
\renewcommand{\arraystretch}{.6}
\tablehead{
\colhead{Configuration} &  
\colhead{Mean $N_p$} &
\colhead{Mean $M_{tot} (\mearth)$} &
\colhead{Median $AMD$} &
\colhead{Median$WMF_\oplus$} &
\colhead{Median $T_{form,\oplus}$(Myr)}}
\startdata
CJS1 & 2.75 & 2.32 & 0.017 & $3.2\times 10^{-3}$ & 140\\
CJS15& 3.5  & 2.77 & 0.003 & $5.7\times 10^{-3}$ & 104\\
\\
EJS1 & 2.5 & 1.55 & 0.010 & $6.3\times 10^{-3}$ & 169\\
EJS15& 3.0 & 2.06 & 0.005 & $2.6\times 10^{-4}$ & 66
\enddata
\tablenotetext{1}{Recall that the CJS1 and EJS1 sims had disks with $r^{-1}$
surface density profiles, while CJS15, EJS15, and all our other simulations
had $r^{-1.5}$ surface density profiles.}
\end{deluxetable*}
\normalsize

\begin{figure}
\centerline{\epsscale{1.}\plotone{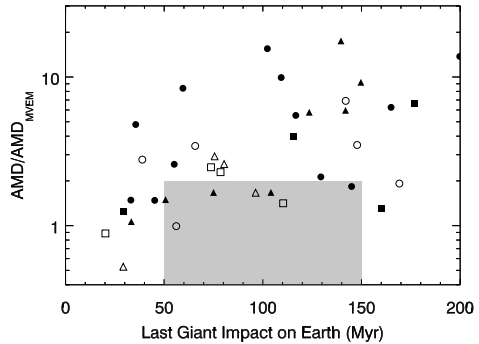}}
\caption{The system angular momentum deficit, normalized to the MVEM value of
0.0018, as a function of the time of the last giant impact for the Earth
analog in each of our simulations.  The symbol for each simulation is the same
as in Fig.~\ref{fig:coll_all}.  The region of successful outcomes is shaded. }
\label{fig:tlg-amd}
\end{figure}

The higher $AMD$ values for $r^{-1}$ simulations appears to be linked to the
mean formation timescale.  Indeed, Figure~\ref{fig:tlg-amd} shows a weak
correlation between the timescale for the last giant impact on Earth
$T_{form,\oplus}$ and the $AMD$ of the system for all 40 simulations.  For
$T_{form,\oplus} <$ 100 Myr, the median $AMD$ is 0.003 and for
$T_{form,\oplus} >$ 100 Myr, the median $AMD$ is 0.010.  We attribute this
trend to the fact that the planetesimal population decays with time and our
simulations have limited resolution.  So for late giant impacts or scattering
events among the embryos, there are fewer planetesimals around to re-damp the
planets by dynamical friction if the planets form more slowly.  This problem
could be alleviated with simulations which continuously regenerate
planetesimals from the debris of giant impacts (e.g., Levison \etal 2005),
because in that case the planetesimal population would be sustained for as
long as the giant impacts occur.

For the remainder of our analysis we consider the giant planet configuration
as the only variable.  We therefore combine the CJS1 and CJS15 simulations
into CJS, and the EJS1 and EJS15 cases into EJS.  Given that the failings of
these simulations are generally the same (see Table 2), we do not expect this
to skew our results.

\begin{figure*}
\centerline{\epsscale{0.5}\plotone{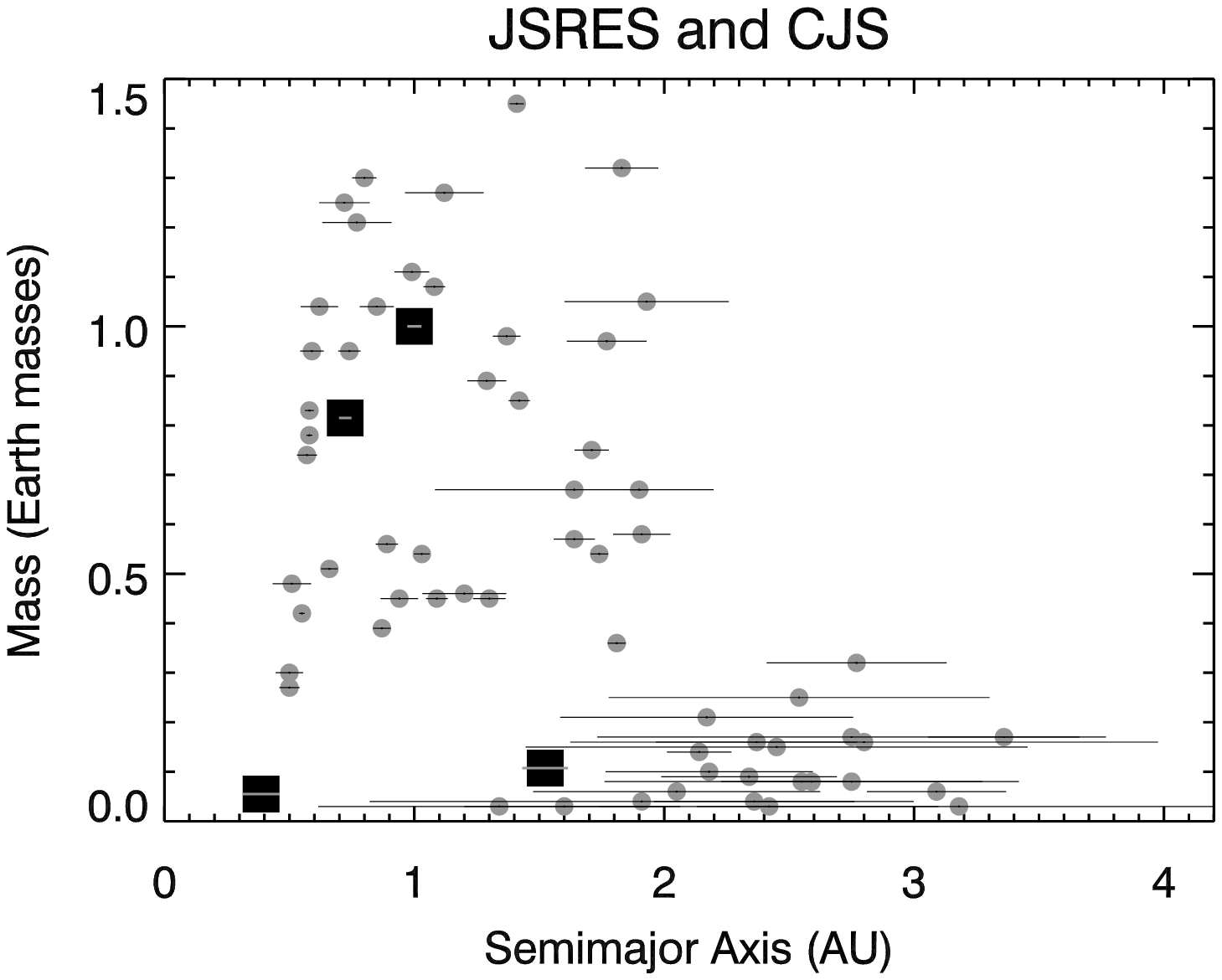}\plotone{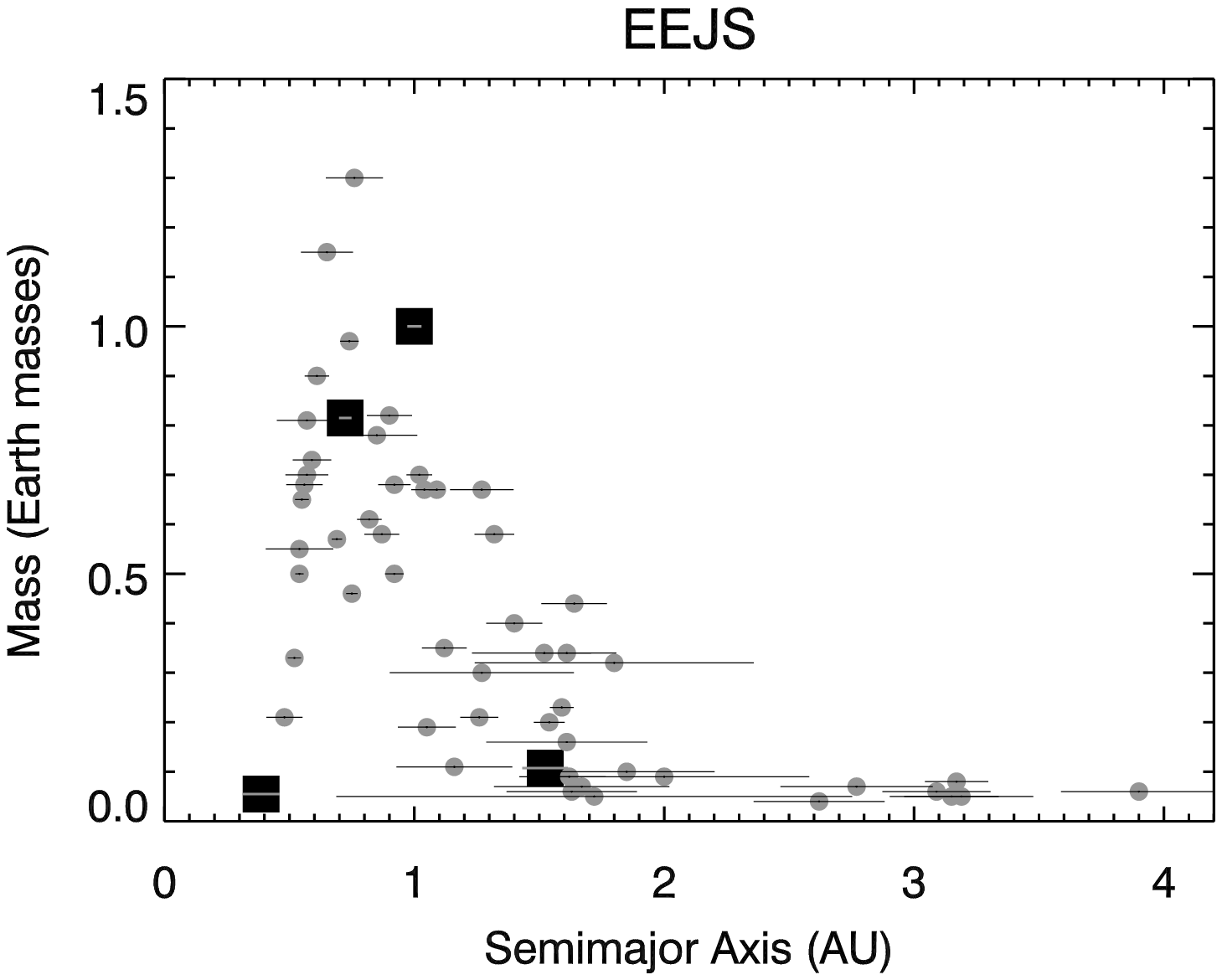}}
\centerline{\epsscale{0.5}\plotone{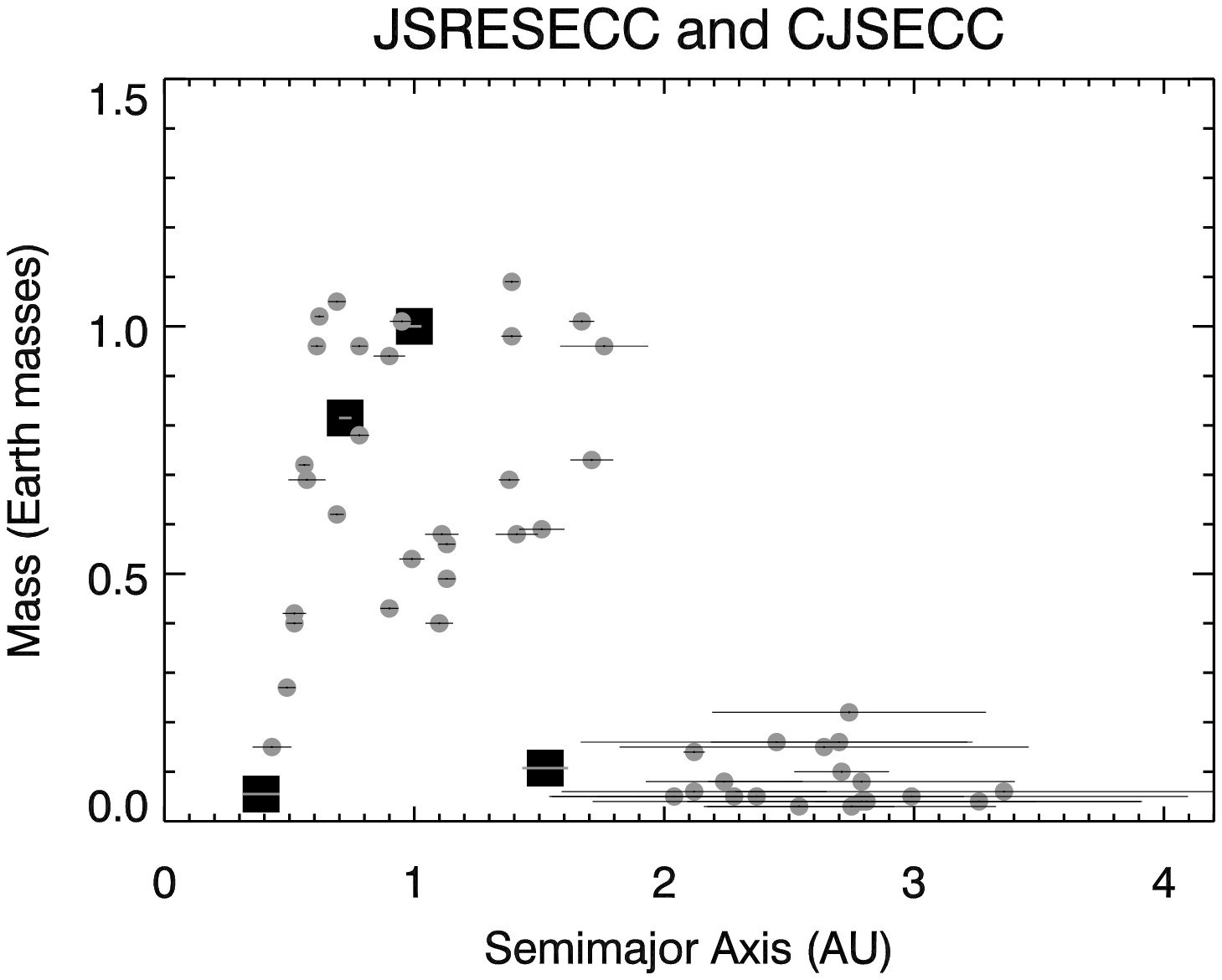}\plotone{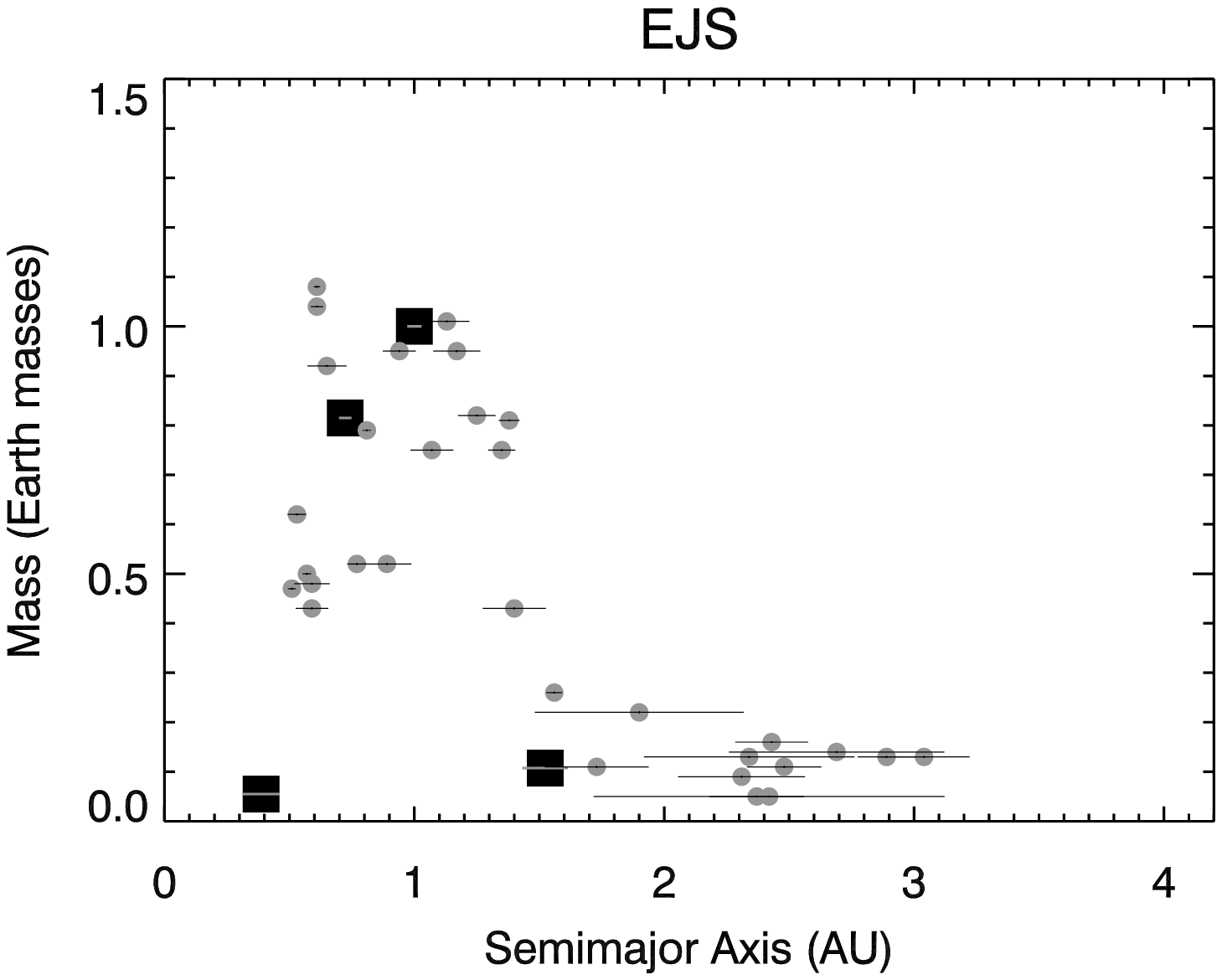}}
\caption{Mass vs. semimajor axis for a range of simulations with different
configurations of Jupiter and Saturn.  Each panel shows all planets that
formed in the relevant simulations (see Table 2) as grey circles, with
horizontal lines representing the orbital eccentricity.  The Solar System's
terrestrial planets are shows as the black squares, with 3 Myr averages for
their eccentricities in grey (taken from Quinn \etal 1991).  }
\label{fig:m-a}
\end{figure*}

Figure~\ref{fig:m-a} shows the mass-semimajor axis distribution of all 40 of
our simulations, grouped into categories with similar giant planet
configurations.  When compared with the Solar System's terrestrial planets, it
is immediately evident that all cases with circular or lower-eccentricity
giant planets fail miserably at reproducing Mars' small size.  Indeed, for the
CJS, CJSECC, JSRES and JSRESECC simulations, planets in Mars' vicinity are
typically 0.5-1 $\mearth$.  The radial distribution of massive planets is much
broader in these cases than in the Solar System, and Earth-sized planets are
commonly formed all the way out to 2 AU.  In contrast, the EJS and especially
the EEJS simulations did a much better job of reproducing Mars' small size.
The EJS simulations have smaller Mars analogs than the CJS and JSRES cases but
in most cases $M_{Mars} \approx 0.3 \mearth$.  In 5/12 EEJS simulations the
Mars analog was between 0.06 and 0.2 $\mearth$, and in each of those 5 cases
the last giant impact occurred before 10 Myr.  The radial mass distributions
for the EJS and EEJS simulations are peaked, as is the case for MVEM.  For the
EJS simulations the peak is close to (or perhaps slightly interior to) 1 AU,
but for the EEJS simulations the most massive planets tend to lie interior to
1 AU and planets at 1 AU are typically half an Earth mass.  This can be
explained as a byproduct of the excitation of the planetesimals and embryos by
the giant planets: planetesimals and embryos on eccentric orbits are most
likely to collide close to perihelion, such that systems with eccentric giant
planets tend to have the most massive planets closer to their stars than for
systems with low-mass or low-eccentricity giant planets (Levison \& Agnor
2003).  

It is also clear from Fig.~\ref{fig:m-a} that simulations with circular or
low-eccentricity giant planets (CJS, CJSECC, JSRES, JSRESECC) tend to strand
massive embryos in the asteroid belt.  These embryos are typically 0.05-0.2
$\mearth$ and would certainly disrupt the observed asteroid distribution.  In
contrast, the EJS and EEJS simulations leave fewer embryos in the asteroid
belt, and those that are stranded are typically smaller.

A trend that is less evident from Fig.~\ref{fig:m-a} is that the total
terrestrial planet mass decreases with the giant planet eccentricity in almost
all cases.  The one exception to this rule is for the JSRESECC simulations,
which have roughly the same total planet mass as the JSRES cases; however, at
the end of each JSRESECC simulation $e_J < 0.01$, so the difference between
the two giant planet configurations is actually fairly minor.  The reason for
the correlation between increased giant planet eccentricity and decreased
total mass in terrestrial planets is simply that eccentric giant planets
perturb terrestrial and asteroidal bodies more strongly and destroy a larger
fraction of the disk via ejection and collisions with the Sun than circular
giant planets (Chambers \& Cassen 2002; Levison \& Agnor 2003; Raymond \etal
2004; O'Brien \etal 2006).  In addition, terrestrial planets can't form as
close to eccentric giant planets as they can to circular giant planets
(Raymond 2006); this may explain the reduced number of stranded asteroidal
embryos for the EJS and EEJS simulations.

\begin{figure}
\centerline{\epsscale{1.}\plotone{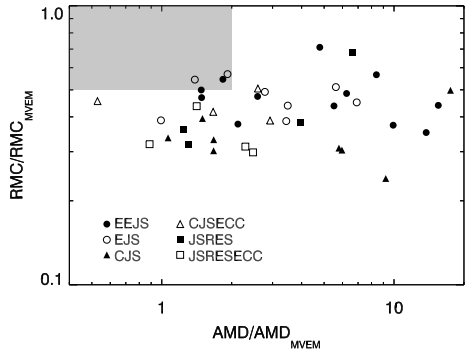}}
\caption{The angular momentum deficit $AMD$ (Eqn. 2) vs. radial mass
concentration statistic $RMC$ (Eqn. 1) for all of our simulations.  Both the
$AMD$ and $RMC$ are normalized to the MVEM values of 0.0018 ($AMD$) and 89.9
($RMC$).  The terrestrial planet system from each simulation is represented as
a single point, following the label at the bottom of the plot (e.g., EEJS
simulations are filled circles, etc).  The region of successful outcomes is
shaded.}
\label{fig:amd-sc}
\end{figure}

Figure~\ref{fig:amd-sc} shows the radial mass concentration statistic $RMC$ as
a function of the angular momentum deficit $AMD$ for the terrestrial planet
system that formed in each simulation.  These statistics are normalized with
respect to the MVEM values of 0.0018 and 89.9.  The systems that formed have a
wide range in $AMD$, from 0.5 to almost 20 times the MVEM value.  In contrast,
systems are clumped in $RMC$ between 0.3 and 0.7 times the MVEM value; none
has $RMC$ higher than 0.71 -- this is similar to the results of Chambers
(2001). Most systems have $AMD$ values somewhat larger than MVEM, although a
few cases have $AMD$ smaller than MVEM (see Table 2). 

Table 4 lists the median $AMD$ and $RMC$ values for each giant planet
configuration.  The $AMD$ varies significantly for the different
configurations, and is smallest for the JSRESECC and CJSECC simulations.  This
is consistent with the results of O'Brien \etal (2006), who formed lower-$AMD$
systems for cases with moderately eccentric giant planets (analogous to our
EJS simulations).  The most eccentric planets formed in the CJS and EEJS
simulations.  The CJS systems were eccentric because the timescale for
clearing out of the asteroid belt and late encounters with remnant embryos was
relatively long such that few planetesimals remained for damping after late
encounters, as discussed above.  The EEJS systems were eccentric mainly
because of the excitation caused by the $\nu_5$ and $\nu_6$ resonances, as
well as direct perturbations by the giant planets.  However, our numerical
resolution also played a significant role: the four EEJS simulations with 1000
planetesimals (EEJS 1-4) had a median $AMD$ of 0.015, but the four simulations
with 2000 planetesimals and slightly less eccentric giant planets (EEJS 9-12)
had a median $AMD$ of 0.0033.  Note, however, that the EEJS terrestrial planet
systems with 2000 planetesimals and $e_J = e_S = 0.1$ (EEJS 5-8) were even
slightly more eccentric than the cases with 1000 planetesimals.  This large
variation in $AMD$ for the EEJS cases is again linked to the presence or
absence of damping at the time of the last encounters between embryos.
Indeed, the mean formation timescale for Earth analogs again scales inversely
with the $AMD$: for EEJS 1-4, 5-8, and 9-12 the median formation timescales
for Earth were 109, 129, and 55 Myr.  This continues an important trend that
we see: longer formation timescales lead to higher $AMD$ because fewer
planetesimals exist for dynamical friction at later times.  This trend is
caused in part by our numerical resolution (1000-2000 planetesimals instead of
billions) and in part because we do not account for impact debris.  However,
the planetesimal population certainly does contribute to decreasing
eccentricities so we believe that this effect is real, although accounting for
other factors should weaken the correlation.

\begin{deluxetable*}{l|ccccc}
\tablewidth{0pt}
\tablecaption{Statistical values of terrestrial planet systems for different
giant planet configurations}
\renewcommand{\arraystretch}{.6}
\tablehead{
\colhead{Configuration} &  
\colhead{Mean $N_p$} &
\colhead{Median $AMD$} &
\colhead{Median $RMC$} &
\colhead{Median $WMF_\oplus$} &
\colhead{Median $T_{form,\oplus}$(Myr)}}
\startdata
CJS & 3.13 & 0.010 & 29.8  & $5.3\times 10^{-3}$ & 123.5 \\
CJSECC & 3.5 & 0.0047 & 40.9  & $1.2\times 10^{-3}$ & 80.4\\
EJS & 2.75 & 0.0062 & 44.2 & $3.3\times 10^{-4}$ & 147.8\\
EEJS\tablenotemark{1} & 3.58 & 0.0099 & 42.6 & $1.1\times 10^{-4}$ & 109.3\\
JSRES & 4 & 0.0071 & 34.2 & $7.5\times 10^{-3}$ & 160.1\\
JSRESECC & 3.75 & 0.0041 & 28.7 & $1.1\times 10^{-3}$ & 78.6\\
\\
Solar System & 4 & 0.0018 & 89.9 & $\sim 10^{-3}$ & 50-150
\enddata
\tablenotetext{1}{Note that there were 12 EEJS simulations, including 8 with
2000 planetesimals.  For the 4 EEJS simulations with 1000 planetesimals (EEJS
1-4), the median $N$, $AMD$, $RMC$, $WMF_\oplus$ and $T_{form,\oplus}$ were
3, 0.015, 50.9, $8.1 \times 10^{-5}$, and 109.3 Myr, respectively.}
\end{deluxetable*}

We performed a suite of two-sided Kolmogorov-Smirnov and Wilcoxon tests to
determine which differences in $AMD$ and $RMC$ in our simulations were
statistically significant.  At the 0.05 level (i.e., $p < 0.05$ where $p$ is
the probability that the two distributions were drawn from the same sample),
we only found differences between the EEJS normal-resolution simulations (EEJS
1-4 = EEJSnr [with 1000 planetesimals], EEJS 5-12 = EEJShr [2000
planetesimals]) and CJSECC, EJS15, and JSRESECC, all with $p < 0.029$ from
Wilcoxon tests.  For the $RMC$ values, the following sets of simulations
provided significantly different values ($p < 0.05$): CJS15 vs. EJS1 ($p <
0.029$), CJS15 vs. EEJShr ($p < 0.016$), EEJShr vs. JSRESECC ($p < 0.016$),
and EJS1 vs. JSRESECC ($p < 0.029$).  If we only include variations of giant
planet configuration and ignore changes in disk surface density profile (for
the CJS and EJS simulations) and resolution (for the EEJS simulations), there
are no statistically significant differences in $AMD$, but for $RMC$ there are
differences between the following configurations: CJS vs. CJSECC ($p <
0.048$), CJS vs. EJS ($p < 0.007$), CJS vs. EEJS ($p < 0.005$), EJS
vs. JSRESECC ($p < 0.016$), and EEJS vs. JSRESECC ($p < 0.008$).  Thus, the
normal-resolution EEJS simulations have significantly higher $AMD$ values than
the other simulations, but increasing the resolution brings them into
agreement with the other cases (just as increasing the resolution for the
other cases would likely also decrease their $AMD$ values).  The CJS
(especially CJS15) and JSRESECC simulations represent the statistically
smallest $RMC$ values of our sample, and the highest come from the $EJS$ and
$EEJS$ simulations, although these are still far below the MVEM value.

\begin{figure}
\centerline{\epsscale{1.}\plotone{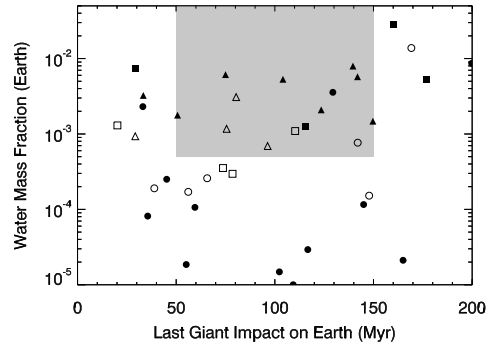}}
\caption{The water content by mass $WMF_\oplus$ vs. the time of the last giant
impact for the Earth analog in each of our simulations.  The symbol for each
simulation is the same as in Fig.~\ref{fig:amd-sc}.  Also as in
Fig.~\ref{fig:amd-sc}, the region of successful outcomes is shaded. }
\label{fig:tlg-wmf}
\end{figure}

We follow Thommes \etal (2008) and define a successful outcome as a system for
which both the $AMD$ and $RMC$ are within a factor of two of the MVEM values.
This successful area is shaded in Fig.~\ref{fig:amd-sc}.  Only four simulations
were successful in terms of $AMD$ and $RMC$, and all had relatively eccentric
giant planets: EJS1-4, EJS15-2, EEJS-11 and EEJS-12.  

Figure~\ref{fig:tlg-wmf} shows the water content by mass $WMF_\oplus$ of the
Earth analog in each simulation vs. the time of the last giant impact on that
same planet $T_{form,\oplus}$.  There is a wide range in both of these
parameters, and some correlation with the giant planet configuration (see
Table 3).  As expected, the majority of dry planets come from the EJS and EEJS
samples.

We define a successful outcome in Fig.~\ref{fig:tlg-wmf} to have
$T_{form,\oplus}$ = 50-150 Myr $WMF_\oplus > 5 \times 10^{-4}$.  The water
constraint requires two oceans of water to have been accreted by the planet,
because Earth's minimum bulk water content is two oceans, one on the surface
and one in the mantle (Lecuyer \etal 1998; 1 ocean = $1.5 \times 10^{24} g$ is
the amount of water on Earth's surface).  Of our 40 simulations, 14 were
successful in $WMF_\oplus - T_{form,\oplus}$ space, but there was no
overlap with the 4 successful cases from $RMC-AMD$ space.  The CJS and CJSECC
cases were the most successful in this respect: 10/12 CJS and CJSECC
simulations satisfied both the $WMF_\oplus$ and $T_{form,\oplus}$
constraints.

\section{Discussion and Conclusions}

In this section we first discuss the degree to which each giant planet
configuration was able to reproduce our observed constraints (\S 6.1).  Next,
we discuss the context of each giant planet configuration in terms of the
Solar System as a whole (\S 6.2).  We then point out the limitations of our
simulations, and plans for future work (\S 6.3). 

\subsection{Success of giant planet configurations in satisfying our
constraints}

Let us quantitatively evaluate how the simulations fared at reproducing our
five constraints using relatively generous values: 1) $M_{Mars} < 0.3
\mearth$, 2) $AMD < 0.0036$ (twice the MVEM value), 3) 50 Myr $<
T_{form,\oplus} <$ 150 Myr, 4) Less than 0.05 $\mearth$ in embryos is stranded
in the asteroid belt, and 5) $WMF_\oplus > 5\times 10^{-4}$.  No single
simulation reproduced all five constraints.  Three simulations reproduced four
constraints. The simulation CJS15-3 reproduced four constraints, but formed a
0.98 $\mearth$ Mars analog at 1.37 AU and a 0.58 $\mearth$ planet at 1.91 AU.
The simulation EJS1-3 formed a small Mars, a wet Earth on the correct
timescale, and stranded one, 0.048 $\mearth$ embryo in the asteroid belt, but
the system's $AMD$ is far too large (0.012).  The simulation EEJS-9 formed a
small Mars, low-$AMD$ planets and a wet Earth with one 0.04 $\mearth$ embryo
in the main belt.  Many simulations reproduced three constraints, but rarely
while forming a small Mars.  As expected from previous work, Mars' small size
was the most difficult constraint to reproduce, and no simulations outside of
the EJS and EEJS configurations form a single Mars analog less massive than
0.5 $\mearth$.

Table 5 crudely summarizes the outcomes of our simulations in terms of the
likelihood of a system with a given giant planet configuration's ability to
quantitatively reproduce our observed constraints using the values above.  We
completed the table as follows.  A configuration is said to reproduce a given
constraint (and receives a ``$\checkmark$'') if at least half of the
simulations were successful for that constraint, using the constraints listed
immediately above.  A configuration is unsuccessful at reproducing a
constraint (and receives a ``$\times$'') if no simulations are successful.  If
isolated cases or a small fraction of simulations are successful, then the
configuration receives a maybe (``$\sim$'').  In one case we bent these rules;
5/12 EEJS simulations formed a Mars analog smaller than 0.3 $\mearth$ (4 cases
$< 0.2 \mearth$) so we gave EEJS a $\checkmark$ for this constraint despite a
slightly less than 50\% success rate.

The most successful giant planet configuration was EEJS (``Extra-Eccentric
Jupiter and Saturn'').  The EEJS simulations reliably satisfied three of our
constraints with two maybes (see Table 5), and the ensemble of EEJS
simulations satisfied all five constraints, although no single simulation did
so.  The EEJS cases reliably formed reasonable Mars analogs in terms of Mars'
mass, orbit, and formation timescale.  Planetary eccentricities were too large
in most cases, but increasing the number of planetesimals (runs EEJS 5-12)
decreased the $AMD$ to close to the MVEM value, and even higher resolution
simulations would presumably continue decrease the $AMD$ to the MVEM value.
The Earth analog formed on the correct, 50-150 Myr timescale in most EEJS
simulations but was too dry in all but three cases.  Almost half (5/12) of the
EEJS simulations finished with no embryos in the asteroid belt, and for the
cases with trapped asteroidal embryos they were typically low-mass.  However,
we note that for the EJS and EEJS configurations the survival of any embryos
in the asteroid belt constitutes a failure because Jupiter and Saturn's orbits
do not allow for any migration, which would be necessary to clear remaining
embryos from the belt -- this issue is discussed further in \S 6.2.

\scriptsize
\begin{deluxetable}{l|ccccc}
\tablewidth{0pt}
\tablecaption{Summary of the success of Jupiter-Saturn configurations for
reproducing inner Solar System Constraints\tablenotemark{1}}
\renewcommand{\arraystretch}{.6}
\tablehead{
\colhead{Config.} &  
\colhead{AMD} &
\colhead{$M_{Mars}$} &
\colhead{$T_{form}$} &
\colhead{Ast. Belt} &
\colhead{$WMF_\oplus$}}
\startdata

CJS & $\checkmark$ & $\times$ & $\checkmark$ & $\sim$ & $\checkmark$ \\

CJSECC & $\checkmark$ & $\times$ & $\checkmark$ & $\times$ & $\checkmark$ \\

EJS & $\checkmark$ & $\sim$ & $\checkmark$ & $\sim$ & $\sim$\\

EEJS & $\checkmark$ & $\checkmark$ & $\sim$ & $\checkmark$ & $\sim$   \\

JSRES &  $\checkmark$ & $\times$ & $\sim$ & $\times$ & $\checkmark$\\

JSRESECC & $\checkmark$ & $\times$ & $\checkmark$ & $\times$ & $\checkmark$

\enddata
\tablenotetext{1}{For each configuration of Jupiter and Saturn, a check
(``$\checkmark$'') represents success in reproducing a given constraint in at
least half the simulations, a cross (``$\times$'') represents a failure to
reproduce the constraint in any simulations, and a twiddle sign (``$\sim$'')
represents a ``maybe'', meaning success in reproducing the constraints in a
smaller fraction of cases.}
\end{deluxetable}
\normalsize

The main reason for the success of the EEJS simulations was the strength of
the $\nu_6$ secular resonance located at $\sim$ 2 AU, which created an
``edge'' to the inner disk, effectively separating it from the asteroid
region, removing material that approached 2 AU and thereby helping to form a
small Mars.  The high eccentricities of Jupiter and Saturn ($e_{Jup,Sat} =
0.07-0.10$) were responsible for the strength of the resonance; the EJS
simulations had $\nu_6$ in the same place but it was too weak to clear out
enough material to form a small Mars.  However, having the $\nu_6$ at this
location also makes it difficult for water-bearing asteroidal material to
enter the inner Solar System and be accreted by the terrestrial planets.

In several of the giant planet configurations that we considered -- CJS,
CJSECC, JSRES, and JSRESECC -- Jupiter and Saturn had lower eccentricities
than their current values.  For all of these simulations, the Earth generally
formed on the correct timescale and the terrestrial planets were
low-eccentricity and water-rich.  However, not a single simulation from these
four cases was able to reproduce Mars' small size, and most simulations
stranded one or more large embryos in the asteroid belt.  Therefore, this work
suggests that low-eccentricity configurations of Jupiter and Saturn cannot
explain the terrestrial planets, in particular Mars' small size, in the
context of our simulations.

It is interesting that none of our simulations was able to reproduce the large
radial mass concentration seen in the Solar System's terrestrial planets ($RMC
= 89.9$ for MVEM vs. 30-50 for most simulations; see Fig.~\ref{fig:amd-sc} and
Eqn 1).  This concentration comes from the large masses and proximity of Venus
and Earth, and the small masses of Mercury and Mars.  Given the difficulty in
producing Mars analogs, it is not surprising that simulations with
low-eccentricity giant planets yield small $RMC$ values.  However, the EEJS
and EJS simulations also yielded $RMC$ values far smaller than MVEM, although
larger than for the other giant planet configurations.  The origin of this
discrepancy is not clear.  It could be related to the structure of the
planetesimal disk; observations suggest that inner, dust-free cavities exist
in many disks around young stars with varying radii, from $< 0.1$ to $\sim$ 1
AU. (e.g., Eisner \etal 2005, Millan-Gabet \etal 2007).  If the Solar Nebula
had a large inner cavity then the inner boundary for the planetesimal disk
could have been at roughly Venus' orbital distance such that the radial
compression of MVEM is a result of accretion in a radially compressed
planetesimal disk.  Alternately, resonant sweeping or tidally-induced
migration from interactions with the residual gas disk or perhaps collisional
debris could compress the terrestrial planet system and increase the $RMC$
(see Thommes \etal 2008).  However, a difficulty with this model is that
Earth's formation timescale is much longer than the typical gas disk lifetime
(see \S 6.3).  

\subsection{Putting the giant planet configurations in the context of the
Solar System}

Our giant planet configurations, described in \S 3.1, represent different
assumptions about the early evolution of the Solar System.  Four of our cases
-- CJS, CJSECC, JSRES, and JSRESECC -- are based on the Nice model, which
requires Jupiter and Saturn to have formed interior to their mutual 2:1 mean
motion resonance (Tsiganis \etal 2005; Morbidelli \etal 2007).  These four
cases assume that delayed, planetesimal scattering-driven migration spread out
the giant planet system, and that the 2:1 resonance crossing of Jupiter and
Saturn triggered the late heavy bombardment (Gomes \etal 2005; see also Strom
\etal 2005).  The other two giant planet configurations -- EJS and EEJS --
have Jupiter and Saturn at their current semimajor axes, meaning that
planetesimal-scattering driven migration is not permitted because this
inevitably leads to spreading out of the giant planets' orbits (Fernandez \&
Ip 1984).  Therefore, the EJS and EEJS simulations assume that the Nice model
is incorrect and that the late heavy bombardment was caused by another
mechanism.  The only other currently-viable such mechanism is the ``planet V''
theory of Chambers (2007) which invokes the formation and delayed instability
of a fifth, sub-Mars mass terrestrial planet at $\sim$ 2 AU.  Thus, for EEJS
and EJS simulations to be consistent with the Solar System's observed history
they must form a planet V with the right mass ($\lesssim M_{Mars}$) and in the
right location ($\sim 2$ AU).

Our results clearly favor the EEJS simulations over giant planet
configurations that are consistent with the Nice model, mainly because of the
EEJS simulations' ability to form Mars analogs.  In addition, the constraints
that were poorly reproduced by the EEJS simulations were in some sense our
weakest, because higher resolution simulations with more planetesimals tend to
lower the $AMD$, and alternate models exist for water delivery to Earth (e.g.,
Ikoma \& Genda 2006; Muralidharan \etal 2008).  Moreover, EEJS simulations 5,
6 and 11 formed reasonable planet V analogs (see Table 2).  In one of these
cases, simulation EEJS-11, a good Mars analog formed at 1.62 AU and a 0.09
$\mearth$ planet V formed at 2.0 AU.  Given the significant inclination of the
planet V in this case (12$^\circ$), the instability timescale could be several
hundred Myr (e.g., Morbidelli \etal 2001).  That simulation therefore may
represent the most self-consistent simulation in our sample, at least when only considering the inner Solar System constraints.

Given that planets are thought to form on circular orbits (e.g., Pollack \etal
1996), what would be the source of the significant eccentricities of Jupiter
and Saturn in the EEJS configuration?  Goldreich \& Sari (2003) proposed that
eccentricities of Jupiter-mass planets could be excited by lindblad resonances
in gaseous protoplanetary disks, but with eccentricities limited by the width
of the gap carved out of the disk by the planet.  D'Angelo \etal (2006) found
that planets embedded in eccentric disks can have their eccentricities
increased to $\sim 0.1$.  However, other studies have shown that eccentricity
growth via planet-disk interactions can only occur for very massive, $>10
M_{Jup}$ planets (Papaloizou \etal 2001; Kley \& Dirksen 2006).  Thus, the
origin of EEJS-type configurations of Jupiter and Saturn remains controversial
but certainly within the realm of current thinking.

Another study has shown that the EEJS configuration may be advantageous for
reproducing the inner Solar System.  Thommes \etal (2008) also took advantage
of strong secular resonances to reproduce the terrestrial planets.  In their
model, collisions between embryos were induced by the inward sweeping of the
$\nu_5$ and $\nu_6$, which occurred as the Solar Nebula dissipates and changes
the gravitational potential (Heppenheimer 1980; Ward 1981).  Their source of
damping is tidal interaction with the gas (Ward 1993; Cresswell \etal
2007). They manage to reproduce several aspects of the terrestrial planets,
including Mars' small size.  However, we note that Thommes \etal (2008)
assumed that the inner disk only contained embryos out to 3 AU and no
planetesimals.

Given these successes, it is tempting to regard the EEJS configuration as the
true configuration of Jupiter and Saturn early in Solar System history.  EEJS
is indeed consistent with the inner Solar System.  In the EEJS scenario,
Jupiter and Saturn must have acquired eccentricities of $\sim 0.1$ at an early
stage, within 1 Myr or so after their formation.  Scattering of unstable
planetesimals and embryos over the $\sim 10^8$ years of terrestrial accretion
decreased their eccentricities to their current values of $\sim 0.05$.
Indeed, the final time-averaged values of $e_{Jup}$ and $e_{Sat}$ in our
simulations are 0.03-0.06 and 0.06-0.10, respectively, very close to their
current orbits.  In this model, Earth's water was delivered in part from
hydrated asteroidal material, but mainly from adsorption of small silicate
grains (Muralidharan \etal 2008), cometary impacts (Owen \& Bar-nun 1995) or
oxidation of a H-rich primitive atmosphere (Ikoma \& Genda 2006).  The late
heavy bombardment of the terrestrial planets can then be explained by the
delayed destabilization of planet V (Chambers 2007), analogs of which did
indeed form in several EEJS simulations.

The outer Solar System provides a strong argument against the EEJS
configuration.  The resonant structure of the Kuiper Belt requires outward
migration of Neptune (Malhotra 1993, 1995; Levison \& Morbidelli 2003).  It is
thought that Neptune migrated outward because of the back-reaction from the
scattering of a many Earth masses worth of remnant planetesimals.  Given that
Neptune cannot easily eject these planetesimals (e.g., Duncan \etal 1987), the
scattering of these small bodies also causes Uranus and Saturn to migrate
outward, and Jupiter, which does eject most of the small bodies, to migrate
inward (Fernandez \& Ip 1984; Hahn \& Malhotra 1999).  This planetesimal
scattering is thought to have started during planet formation and lasted for
$10^7 - 10^9$ years, with more rapid migration corresponding to a shorter
migration, a more massive planetesimal disk or closer proximity between
Neptune and the disk's inner edge (Gomes \etal 2004, 2005).  Excess depletion
just exterior to strong resonances in the asteroid belt provides empirical
corroboration of giant planet migration (Minton \& Malhotra 2009), although
the migration timescale cannot be constrained.  Thus, the current orbits of
Jupiter and Saturn are not thought to be their orbits at the time of
formation, and it is their orbits at early times that affected terrestrial
planet formation.  In particular, Jupiter and Saturn must have formed in a
more compact configuration, because scattering-induced migration always causes
their orbits to diverge.

Is it possible to reconcile the EJS and EEJS configurations with the outward
migration of Neptune that is required by current Kuiper Belt models?  If a
small Mars formed because of the $\nu_6$ resonance, then the $\nu_6$ must have
been in roughly its current location during Mars' formation.  Hf/W isotopes
suggest that Mars formed much faster than Earth, and did not undergo any giant
impacts after 1-10 Myr (Nimmo \& Kleine 2007).  In addition, the location of
the $\nu_6$ is highly sensitive to Jupiter and Saturn's semimajor axes such
that a more compact configuration places the $\nu_6$ in the asteroid belt,
farther from Mars and less likely to influence its growth (see Fig. 2 of
Minton \& Malhotra 2009).  Migration of Jupiter and Saturn must therefore have
been completely finished in a few Myr at the latest to form a small Mars.
Would such an early migration fit our current understanding?  At such early
times there likely existed a population of embryos in the asteroid belt
(Wetherill 1992).  If Jupiter and Saturn's migration occurred that early, then
embryos would probably have smeared out the observed depletion of asteroids
exterior to asteroid belt resonances (Minton \& Malhotra 2009). In addition,
this very rapid migration must also have taken place in the presence of some
amount of residual disk gas, which would certainly have affected the
scattering dynamics (Capobianco \etal 2008), and would likely have been far
too efficient in trapping Kuiper Belt objects in resonances with Neptune
(e.g., Mandell \etal 2007).  This suggests that the EEJS configuration is at
best marginally consistent with the outward migration of Neptune needed by
models of the Kuiper Belt.  Neptune's migration would have to have occurred
early and been very rapid.  In addition, to remain consistent with the late
heavy bombardment, a fifth terrestrial planet must have formed at $\sim$ 2
AU. (Chambers 2007).

An alternate hypothesis for the evolution of the outer Solar System suggests
that there initially existed several additional ice giant planets (Chiang
\etal 2007).  Could such a scenario be consistent with the EEJS configuration?
In this model, instabilities among the ice giants would have led to the
ejection of excess ice giants and dynamical friction could re-circularize the
orbits of Uranus and Neptune (Ford \& Chiang 2007).  However, in order to
avoid disrupting the EEJS configuration by injecting planetesimals into the
Jupiter-Saturn region, those planetesimals would need to have been dynamically
ejected by the ice giants themselves, which is unlikely given their relatively
small masses.  In addition, a detailed simulation of this model suggests that
a population of many ice giants would not eject each other but would simply
spread out by interactions with the planetesimal disk (Levison \& Morbidelli
2007).  Thus, we are left with the same issue, that the giant planets would
have had to clear out the outer Solar System in less than Mars' formation
timescale of a few Myr.  This scenario therefore does not appear to be
consistent with the EEJS configuration.

The Oort cloud provides an additional argument against the EEJS configuration.
If the giant planets reached their current orbits very quickly (as required by
EEJS), then the Oort cloud had to form very quickly as well.  To populate the
inner and outer classical Oort cloud, the current galactic environment (or one
moderately denser) is required (Brasser \etal 2006; Kaib \& Quinn 2008).
However, to capture Sedna (Brown \etal 2004), a significantly denser
environment is needed.  These two constraints are not at odds if the giant
planets evolved on long timescales, because the galactic environment can
change on a 10-100 Myr timescale (e.g., Lamers \etal 2005).  However, if the
giant planets finished migrating, and therefore finished clearing the
planetesimal disk, within a few Myr, then the Solar System should have either
an Oort cloud {\em or} Sedna, but not both.

Thus, we are left with a problem: the giant planet configuration which best
reproduces the terrestrial planets (EEJS) is at best marginally consistent
with the current view of the outer Solar System's evolution, and likely
inconsistent.  On the other hand, configurations that are based on current
outer planet evolution models, in particular the Nice model, cannot form a
small Mars, an inescapable inner Solar System constraint.  Given that we do
not have a choice for the configuration of Jupiter and Saturn at early times
that satisfies all of our constraints and is also consistent with the
evolution of the outer Solar System, we cannot reject the Nice Model.

\subsection{Simulation limitations and future work}

It is important to note that our simulations are missing several physical
effects that could be important.  These effects include collisional
fragmentation (Alexander \& Agnor 1998), dynamical effects of collisional
debris (Levison \etal 2005), tidal damping and resonant sweeping from a
residual amount of nebular gas (Kominami \& Ida 2002; Nagasawa \etal 2005;
Thommes \etal 2008).

We do not think that including collisional fragmentation would affect our
results for two reasons: 1) Alexander \& Agnor (1998) saw little difference in
their simulations when they included a simple fragmentation model, and 2) the
velocities of embryo-embryo collisions were slow enough that the majority were
accretionary rather than erosive (Agnor \& Asphaug 2004).

We also assumed that all embryo-planetesimal impacts were accretionary,
although these do tend to occur at larger velocities than embryo-embryo
impacts because of viscous stirring (see Fig.~\ref{fig:coll}).  Melosh (1984)
showed that it is difficult for km-sized impactors to accelerate material to
speeds greater than one third of the impact speed.  Based on this, we can
assume that planetesimal impacts at speeds of less than 3 times the escape
speed $v_{esc}$ are 100\% accretionary; 83\% of the collisions in our
simulations meet this criterion.  Furthermore, only for impacts with
$v/v_{esc} > 7-10$ do small-body collisions actually become erosive (O'Keefe
\& Ahrens 1977; Svetsov 2007).  Only 2 out of 3400 collisions in our
simulations had $v/v_{esc} > 7$.  Thus, our assumption of perfect accretion
for planetesimal-embryo impacts probably does not affect our results if
planetesimals were indeed km-sized.

The dynamical effects of collisional debris are certainly important, although
their effects have barely been explored because of numerical limitations.  A
continuous source of small bodies after large impacts would likely reduce the
eccentricities of the terrestrial planets and make it easy to reproduce the
low $AMD$ of MVEM (Levison \etal 2005).  In addition,
planetesimal-planetesimal collisions create collisional cascades that can
grind planetesimals to dust, which can then be removed from the system (e.g.,
Kenyon \& Bromley 2006).  However, interactions between the gas disk and small
fragments at early times may accelerate embryo growth by stopping collisional
cascades and allowing embryos to accrete small fragments (Kenyon \& Bromley
2009).

Tidal damping from small amounts of remnant disk gas can help reduce the
eccentricities of the terrestrial planets, as shown by Kominami \& Ida (2002,
2004) and Agnor \& Ward (2002).  Gaseous disks around other stars are thought
to disperse in 5 Myr or less (Haisch \etal 2001; Briceno \etal 2001; Pascucci
\etal 2006).  Such lifetimes correspond to a decrease in gas density to a
level of about 10\% of the minimum-mass nebular density, while tidal damping
can operate down to a level of $10^{-3}$ to $10^{-4}$.  Nonetheless, for this
to be important the gas density must be at least $10^{-4}$ for the timescale
of Earth's accretion, 50-150 Myr (Touboul \etal 2007).  For a steady accretion
model, the gas density $\Sigma$ decreases relatively slowly with time $t$, as
$t^{-3/2}$ (Lynden-Bell \& Pringle 1974).  Thus, if the Solar Nebula evolved
smoothly with a characteristic timescale of 1 Myr, then its density would have
decreased to 10\% after 5 Myr, but would still be $10^{-3}$ after 100 Myr.  In
such a scenario, tidal damping would indeed be an important effect on
terrestrial accretion.  However, the final phases of disk dissipation are
thought to occur after a few Myr (Haisch \etal 2001) on a $\sim 10^5$ year
timescale (Simon \& Prato 1995; Wolk \& Walter 1996) and relatively violently,
via photo-evaporation (Hollenbach \etal 1994; Johnstone \etal 1998) or
potentially the MRI instability (Chiang \& Murray-Clay 2008).  Future, more
sensitive observations that probe smaller gas densities in disks around young
stars will shed light on this issue, but our interpretation of the current
state of knowledge is that it is unlikely that tidal damping would
dramatically change our results.  Indeed, we suspect that the most important
source of damping during accretion is likely to be small bodies, i.e.,
planetesimals and collisional debris.

Secular resonance sweeping during the dispersal of the Solar Nebula could play
an important role in terrestrial planet formation if Jupiter and Saturn's
orbits have eccentricities of at least $\sim$ 0.05 (Nagasawa \etal 2005;
Thommes \etal 2008).  However, if the giant planets' orbits are less eccentric
then secular resonances are too weak to have much effect on the outcome
(O'Brien \etal 2007).  We note that the model of Thommes \etal (2008) invokes
this resonant sweeping to induce collisions as well as to shepherd embryos
from the asteroid belt inward.  However, Thommes \etal assume that the entire
inner disk is composed of embryos and do not include any planetesimals in
their simulations.  Given the strength of aerodynamic gas drag on km-sized
bodies (Adachi \etal 1976), planetesimals should also have been shepherded
inward by the resonance (e.g., Raymond \etal 2006b; Mandell \etal
2007). Indeed, the mass distribution in the area swept out by the secular
resonances may have been drastically altered if the resonances were strong
enough.  We are currently exploring the consequences of this idea.

We did not test an exhaustive number of configurations of Jupiter and Saturn.
Indeed, given the relative success of the EEJS configurations, it would be
interesting to explore the resonant behavior for other allowed configurations
such as, for example, Jupiter and Saturn on their CJS orbits but with
eccentricities of 0.1.  We plan to search the parameter space of giant planet
configurations in future work, focusing on systems with strong mean motion and
secular resonances in the inner Solar System.

The majority of this paper has dealt with the effects of the giant planet
configuration on the accretion of the inner Solar System.  However, we note
that the disk's surface density distribution could also play a strong role in
the outcome (Raymond \etal 2005).  Indeed, we did see changes in the 8
simulations we performed with a flatter, $r^{-1}$ surface density profile.  We
note that there exists an alternate model that attributes Mars' small mass
with the Solar Nebula's density structure (Jin \etal 2008).  In Jin \etal's
model the disk is ionized only in certain regions, causing radial variations
in the strength of the magneto-rotational instability and therefore in the
disk viscosity (Balbus \& Hawley 1991).  At the interface between an outer
low-viscosity and an inner high-viscosity regime a local dip in the surface
density can be created.  Jin \etal (2008)'s model has this dip at about Mars'
orbital distance.  This dip is quite deep but very narrow, although it could
have swept over a region of radial with $\sim$1 AU in the lifetime of the
disk.  When considering accretion in such a disk, it is important to note that
the typical planetary feeding zone in our sample has a width of 2-3 AU, much
wider than the widest possible gap created in the disk of Jin \etal.  Indeed,
preliminary simulations using Jin \etal's disk form Mars analogs that are
$\sim 5 \rm M_{Mars}$.

Our simulations are among the highest-resolution to date, but we are still
resolution-limited.  To accurately model Mars' growth we require that the
embryo mass at Mars' orbital distance is smaller than Mars itself, in our case
by a factor of 3-6 (Fig.~\ref{fig:init}).  This mass sets the inter-embryo
spacing and we are subsequently limited by the number of particles we can
integrate in a reasonable time (of several months).  Given these restrictions,
our simulations may not always adequately model dynamical friction, because
the planetesimal-to-embryo mass ratio is relatively small ($\sim$ 10 at Mars'
orbital distance; smaller closer in and larger farther out).  Indeed, the EEJS
simulations with 2000 planetesimals yielded smaller eccentricities than those
with 1000 planetesimals because we could resolve the damping at late times.
Our approach contrasts with that of Chambers (2001) and O'Brien \etal (2006),
who used fixed-mass embryos and therefore had fixed planetesimal-to-embryo
mass ratios of 10 and 40, respectively.  The advantage of their approach is
that dynamical friction is more consistently modeled, but the disadvantage is
that the number of embryos is small and the embryo mass is large, about one
Mars mass.  Despite the limited resolution and the difficulty with dynamical
friction, all of our constraints were met in some simulations, including cases
with $AMD$ values lower than the MVEM value.  We therefore think that our
approach is valid, and we anticipate that faster computers will certainly
improve the outlook for understanding the origin of the inner Solar System in
the coming years.  

\vskip 0.2in

We thank John Chambers and an anonymous second reviewer for comments that
helped improve the paper.  We also thank John Armstrong for his help in
running the bulk of these simulations at Weber State University.  We
acknowledge useful discussions with Hal Levison, Amara Graps, Michel
Dobrijevic, Franck Selsis, Avi Mandell, and Matija Cuk. S.N.R. is grateful for
support from NASA Origins of Solar Systems (grant NNX09AB84G), a NASA
Postdoctoral Program fellowship, and to NASA's Astrobiology Institute via the
Virtual Planetary Laboratory lead team.  D.P.O. acknowledges support from
NASA's Planetary Geology and Geophysics and Outer Planets Research programs.


\end{document}